\documentclass[1p]{elsarticle}

 \biboptions{comma,square}

\usepackage[T1]{fontenc}
\usepackage[latin9]{inputenc}
\usepackage{verbatim}
\usepackage{amstext}
\usepackage{amssymb}
\usepackage{color}
\usepackage{graphicx,psfrag}
 \usepackage{amssymb}

\def\beq{\begin{equation}}
\def\eeq{\end{equation}}
\def\beqa{\begin{eqnarray}}
\def\eeqa{\end{eqnarray}}
\def\beg{\begin{lyxgreyedout}}
\def\eeg{\end{lyxgreyedout}}

\begin{document}

\begin{frontmatter}

\title{Gauge Interaction as Periodicity Modulation}

  \author{Donatello Dolce}
  \ead{ddolce@unimleb.edu.au}

\date{\today}
\address{Centre of Excellence for Particle Physics (CoEPP),\\  The University of Melbourne,\\   Parkville 3010 VIC, \\ Australia.
}

\begin{abstract}
The paper is devoted to a geometrical interpretation of gauge invariance in terms of the formalism of field theory in compact space-time dimensions \cite{Dolce:2009ce}. In this formalism, the kinematic information of an interacting elementary particle is encoded on the relativistic geometrodynamics of the boundary of the theory through local transformations of the underlying space-time coordinates.  
Therefore, gauge interaction is described as invariance of the theory under local deformations of the boundary,
the resulting local variations of field solution are interpreted as internal transformations, and the internal symmetries of the  gauge theory  turn out to be related to corresponding local space-time  symmetries. In the case of local  infinitesimal isometric transformations,  Maxwell's kinematics and  gauge invariance are inferred directly from the variational principle.         
 Furthermore  we explicitly impose periodic conditions at the boundary of the theory as semi-classical quantization condition     
in order to investigate the quantum behavior of gauge interaction. In the abelian case the result is a remarkable formal correspondence with scalar QED.     
\end{abstract}

\end{frontmatter}

%\pagebreak
%\tableofcontents{}
%
%\addcontentsline{toc}{section}{Introduction}
%\pagebreak

\section*{Introduction\label{sec:0}}

In 1918 Weyl \cite{Weyl:1918ib} introduced the idea of gauge invariance in field theory in an attempt to describe electromagnetic interaction as  invariance  under  local transformations of space-time coordinates. 
In particular he tried to extend  the principle of relativity of the choice of reference frame  to the choice of local units of length. For this reason the idea was   named \emph{gauge} invariance. 
 The requirement of gauge invariance  in fact necessitates the introduction of a new field in the theory, named   \emph{compensating field} \cite{Blagojevic:2002du}. It cancels all  the unwanted effects of the local transformations of variables such as the related  \emph{internal transformation}   of the matter fields, and enable the existence of a local symmetry.  Weyl noticed that such a compensating  field has important analogies with the electromagnetic potential. His proposal was very appealing because of its deep analogies with the geometrodynamical\footnote{The term ``geometrodynamics'' is used as a synonym to indicate a geometrical description of interactions \cite{Ohanian:1995uu}} description of General Relativity (GR).  
 Even though the compensating field associated with local Weyl transformations can be successfully  used to represent gravitational interactions, further developments soon showed that such geometrodynamical description of electromagnetism (EM) was not possible.  Contrary to the experimental evidence, the compensating field associated with  Weyl's gauge interacts in the same manner with particle and antiparticles. The possibility of a geometrodynamical description of EM in terms of local transformations of the underlying space-time coordinates  was abandoned, though the terms gauge invariance still remain for historical reasons. In order  to reproduce the correct gauge field (\emph{i.e.} compensating fields with an imaginary unit in front with respect to the original Weyl compensating fields) an ``internal'' symmetry under local phase transformations of the matter fields  was postulated.  
 These  \emph{internal transformations} of the fields originate ordinary gauge interaction.
 
Another important attempt for a geometrodinamical description of gauge interaction 
is represented by the so-called Kaluza's miracle. In \cite{Kaluza:1921tu} Kaluza showed that classical EM  actually has a well defined geometrodynamical interpretation, but this interpretation involves   an eXtra-Dimension (XD) --- at least as a ``mathematical trick''. The gauge field appears  as entries in the space-time components on XD Kaluza's metric and the Maxwell equations are retrieved from the fifth components of the five-dimensional (5D) Einstein equations. This represent  a \emph{de facto} 5D geometrodynamical unification of electromagnetic  and gravitational interaction.
Also to be considered is  Klein's original proposal, which was to impose Periodic Boundary Conditions (PBCs) at the ends of a compact XD (cyclic XD with topology $\mathbb S^{1}$) in the attempt of a semi-classical interpretation of Quantum Mechanics (QM). In
\cite{Klein:1926tv} he actually noticed that such PBCs provide an analogy with the Bohr-Sommerfeld
quantization condition --- in particular he used this hypothetical cyclic XD to interpret  the quantization of the electric charge.   Similarly it is important to mention Wheeler's program of reduction of every physical phenomenon to a purely geometric aspect, as summarized by his slogan ``Physics is Geometry!'' \cite{Misner:1957mt} or the Rainich's ``already unified field theory'' where the (``square'') of the electromagnetic field strength is reinterpreted in terms of the Ricci tensor \cite{Rainich}.    

In this paper we will investigate these attempts for a geometrical description of gauge invariance in terms of the formalism of field theory in compact space-time dimensions (compact 4D). This formalism, defined in \cite{Dolce:2009ce} and summarized in sec.(\ref{Compact4D:formalism}), see also \cite{Dolce:2009cev4,Dolce:2010ij,Dolce:QTRF5,Dolce:Dice,Dolce:FQXi,Dolce:Cyclic}, can be regarded as the natural realization of the de Broglie assumption at the base of wave mechanics (wave-particle duality). 
In fact, by using de Broglie's words, the formalism is based on the fundamental assumption ``\textit{of existence of a certain periodic phenomenon of a yet to be determined character, which is to be attributed to each and every isolated energy parcel [elementary particle]}'' \cite{Broglie:1925,Broglie:1924}.  This so-called ``de Broglie periodic phenomenon''  \cite{1996FoPhL} or ``de Broglie internal clock'' \cite{2008FoPh...38..659C} has been implicitly tested by 80 years of successes of QFT and indirectly observed in a recent experiment \cite{2008FoPh...38..659C}.  We will realize this assumption by imposing  the de Broglie periodicity as a constrain to a free field. That is to say, as the solution of a bosonic action in  compact  4D and   PBCs. This means that the compactification lengths as well as the boundary of the theory, are explicitly related to the de Broglie periodicity. Therefore the resulting field solution will be a sum over harmonic modes similarly to a string vibrating in compact dimensions. That is, the PBCs (allowed by the variational principle) will be used  as quantization condition similarly to the semi-classical quantization of a particle in a box.  As noticed in \cite{Dolce:2009ce} our description can be regarded as the full relativistic generalization of sound waves. A sound source is a vibrating string in compact spatial dimension within a classical framework. Similarly, an elementary particle will be described as a  vibrating string in compact 4D within a relativistic framework. Indeed, as we will mention later, the theory can be regarded as a particularly simple type of string theory. 

In the first part of the paper we will mainly investigate classical gauge invariance, in particular classical EM. 
In par.(\ref{sec:Towards-Interactions}), we will investigate in a geometric  way how the de Broglie spatial and temporal periodicities of an elementary boson,
\emph{i.e.} of a  corresponding single mode of a Klein-Gordon (KG) field, varies dynamically under local transformations of reference frame or interactions. 
The temporal and spatial de Broglie periodicities can be used to describe the four-momentum of a particle, according to the de Broglie phase harmony. They can be represented as a four-vector. Such a de Broglie four-periodicity  is derived by Lorentz transformations from the invariant periodicity of the proper time associated with the mass of the particle. Indeed it describes a ``periodic phenomenon'' of topology  $\mathbb S^{1}$.
According to de Broglie,  the  local and retarded variations of four-momentum (or four-frequency) of a particle occurring during a given relativistic interaction scheme can be equivalently described in terms of local and retarded variations (modulations) of four-periodicity of a corresponding periodic phenomenon.  In  the formalism of field theory in compact 4D the de Broglie four-periodicity is described by  the boundary of the theory. Thus  the kinematics of a particle in a given interaction scheme  turns out to be encoded in corresponding geometrodynamics of the boundary of the theory --- in a manner of a  holographic principle. 
  Furthermore, deformations of the boundary of the action can be obtained from corresponding transformations of the space-time variables, \emph{i.e.}   diffeomorphisms of the compact 4D of the theory. Hence this description of interaction mimics linearized gravity. In fact  gravitational interaction can be described in terms of modulations of periodicity of reference clocks encoded in corresponding geometrodynamics of the underlying 4D. 
  In \cite{Dolce:AdSCFT} we have noticed that such a geometrical description applied to the Quark-Gluon-Plasma (QGP) exponential freeze-out actually provides an interesting parallelism  with phenomenological aspects of the AdS/QCD correspondence.

Thus, similarly to GR, we want to describe interaction  in terms of  invariance of the theory under a corresponding local transformations of variables. In ordinary field theory simple isometric transformations of the underlying space-time dimensions have no effect. This is essentially because in ordinary field theory the BCs have a very marginal role. The KG field used for practical computations  is the most generic solution of the KG equation. However, as evident in our formalism, a generic transformation of variables implies a corresponding  deformation of  the boundary of the theory and in turn, through the PBCs, of the de Broglie periodicity of the field solution of the theory.  
Indeed we have a variation of field solutions even in the case in which the transformation of variables is a simple  isometry. In fact, even if the transformation is from a flat space-time to a flat space-time, leaving the structure of the action invariant, it could involve a corresponding deformation (rotation) of the boundary of the theory. We will show that classical gauge interactions can be derived by requiring invariance of the theory in compact 4D  under local isometries and by applying the variational principle at the boundary. To figure out the idea we can imagine to describe the trembling motion \cite{Hestenes,Sidharth:2008te} (we will use the  german term \emph{zitterbewegung} for the analogy of the idea to the Scr\"odinger's \emph{zitterbewegung} model \cite{Schrodinger}) of a charged particle interacting with an electromagnetic field in terms of local transformations of reference frame.  For this reason, in this paper we limit our study to the approximation of local isometries; that is, particular isomorphisms where the lengths of the four-vectors are in a first approximation preserved (contrary to the Weyl invariance we do not consider  scale transformations). The  case of local scale (conformal) invariance  has been partially investigated in \cite{Dolce:AdSCFT} through the dualism with XD theories.

More in detail, in sec.(\ref{Isometry})  we will interpret the variation of the field solution associated to a local isometry in terms of internal symmetries of the field.   A local isometry induces a minimal substitution formally equivalent to the one of classical EM.    
From the variational principle it is possible to find out that  this description formally reproduces  
 the Noether current of ordinary gauge interactions. The symmetry of the gauge transformation  turns out to be the symmetry of the isometry which originates it. The resulting gauge field describes the local deformation of periodicity of a matter field under a local transformation of variables. 
 
  In sec.(\ref{Gauge:interaction}), we will see the possibility to write fields with different periodicities in an action with persistent boundary by using gauge invariant terms. The requirement of gauge invariance is therefore derived from the variational principle. Gauge transformations allow one to tune the periodicity of the different fields of a theory in order to minimize the action at the common boundary.  For the same reason we will see that only particular types of  local isometries, which we will call  \emph{polarized}, are allowed by  the variational principle. These polarized  local isometries reproduce Maxwell dynamics for the gauge field.
  Thus the geometrodynamics associated with these particular local isometries reproduce formally classical gauge   theory.   
  
 In \cite{Dolce:AdSCFT} we have shown that field theory in compact 4D is dual to the Kaluza-Klein (KK) field theory. Under this dualism the geometrodynamical description of gauge interaction proposed here  can be regarded as a purely 4D formulation of Kaluza's original proposal. 

 The formalism of field theory in compact 4D also provides an interesting analogy with Klein's original proposal. In fact, in  \cite{Dolce:2009ce,Dolce:AdSCFT} we have shown that the PBCs at the geometrical boundary of the theory represent a semi-classical quantization condition. This can be regarded as the relativistic generalization of the quantization of a particle in a box. This idea is inspired to the 't Hooft determinism \cite{'tHooft:2001ar} and the stroboscopic quantization \cite{Elze:2003tb} where QM is interpreted as an emerging phenomenon associated to some underlying cyclic dynamics.  In our theory the Feynman Path Integral (FPI) formulation  arises in a semi-classical way as interference between classical paths with different winding numbers associated with the underlying cyclic geometry $\mathbb S^{1}$ of the compact 4D \cite{Dolce:2009ce,Dolce:2009cev4}.  Moreover the theory has implicit commutation relations. Generalizing this description, in sec.(\ref{Quantum:Formalism}) 
 we will finally find that the modulation of four-periodicity of an interacting  cyclic field  is formally described in local Hilbert spaces by the ordinary Scattering Matrix of QM. The space-time evolution  associated to such local isometries of the compact 4D is formally described by the ordinary FPI of scalar QED. 
 
\section{Compact space-time formalism}\label{Compact4D:formalism}

In  relativistic mechanics  
every isolated elementary system (free elementary particle)
has associated persistent four-momentum $\bar{p}_{\mu}=\{\bar{E}/c,- \mathbf{\bar{p}}\}$.
On the other hand, the de Broglie-Planck formulation of QM prescribes
that to every particle with four-momentum there is  a corresponding ``periodic phenomenon'' with four-angular-frequency $\bar{\omega}_{\mu}=\bar{p}_{\mu}c/\hbar$, \emph{i.e.} with corresponding de Broglie four-periodicity $T^{\mu}=\{T_{t},\vec{\lambda}_{x}/c\}$.
 The topology of this so-called ``de Broglie periodic phenomenon'' is $\mathbb S^{1}$.  In fact it is fully characterized by the proper time periodicity $T_{\tau}$ \cite{Broglie:1925,Broglie:1924}. In a generic frame, the spatial and temporal components of the de Broglie four-periodicity $T^{\mu}$  are obtained through Lorentz transformations: $cT_{\tau} = c \gamma T_{t} - \gamma \vec \beta \cdot \vec \lambda_{x}$. The energy and the momentum of a particle with mass $\bar M$ in the new reference frame is $E= \gamma \bar M c^{2}$ and $\mathbf{\bar{p}} = \gamma \vec \beta \bar M c$, respectively. In the ``de Broglie periodic phenomenon'', also known as ``de Broglie internal clock'', the proper time periodicity is fixed by the  mass of the particle, according to  $T_{\tau} \bar M c^{2} = h$. Therefore, in a generic reference frame,  we have de Broglie-Planck relation (de Broglie phase harmony)
 \begin{equation}
 c {\bar{p}_{\mu}}T^{\mu} = h  \,.
  \label{eq:PdB:relation}\end{equation}
   
Similarly to the de Broglie assumption of ``periodic phenomenon'', in the compact 4D formalism \cite{Dolce:2009ce} we assume that every elementary bosonic particle with four-momentum $\bar{p}_{\mu}$ is described by
 an intrinsically periodic bosonic field with de Broglie four-periodicity $T^{\mu}$ fixed dynamically through PBCs  (\ref{eq:PdB:relation}). 
This means that, as long as we describe  free particles, \emph{i.e.} persistent $\bar{p}_{\mu}$,  the intrinsic four-periodicity $T^{\mu}$ of the fields must be assumed to be persistent.   
Thus we describe such a free bosonic field with persistent four-periodicity $T^{\mu}$ as the field solution of a relativistic bosonic action in compact 4D with PBCs (PBCs are represented by the circle in the volume integral $\oint$)
\begin{equation}
\mathcal{S}=\oint_{0}^{T^{\mu}}d^{4}x\mathcal{L}(\partial_{\mu}\Phi(x),\Phi(x))\,.\label{free:act}
\end{equation}
Roughly speaking, in a ``de Broglie periodic phenomenon'' the whole physical information is
contained in a single four-period $T^{\mu}$,  \cite{Dolce:2009ce}.  In fact, \textit{`` By a clock we understand anything characterized by a
  phenomenon passing periodically through identical phases so that we must
  assume, by the principle of sufficient reason, that all that happens in a
  given period is identical with all that happens in an arbitrary period''}
  A. Einstein \cite {Einstein:1910}. Indeed, under this assumption of intrinsic periodicity, every isolated elementary particle can be regarded as a reference clock. This aspect will play a central role since, in analogy with GR, we will describe interactions in terms of modulations of periodicity of these  ``de Broglie internal clocks''. 

 It is important to note that the PBCs  minimize the above bosonic action at the boundary. That is, every bosonic field (solution of the Euler-Lagrange equations) with four-periodicity $T^{\mu}$  automatically minimizes the above action \cite{Dolce:2009ce}.
This is the  fundamental reason why the theory turns out to be  fully consistence with Special Relativity (SR) \cite{Dolce:2009ce}.
 Indeed PBCs
have the same formal validity of the usual Stationary (or Synchronous) BCs (SBCs) of
 ordinary field theory (\emph{i.e} fixed values of the fields at the boundary).  The Lorentz covariance of relativistic bosonic actions is preserved by PBCs.

The theory satisfies time ordering and relativistic causality.
In fact, just as  Newton's law of inertia does not imply that every point particle
moves on a straight line (persistent $\bar p_\mu$), our assumption of intrinsic periodicities
does not mean that our  field solutions have always persistent periodicities $T^{\mu}$.
Indeed, the four-periodicity $T^{\mu}$ must be regarded  as local and dynamical  as the four-momentum $\bar p_\mu$, according to (\ref{eq:PdB:relation}).  
 The retarded and local variations of four-momentum occurring during interactions imply
 retarded and local variations (modulations) of the intrinsic four-periodicity of the particles. Events in time (\emph{i.e.} interactions) are characterized by variations of periodic regimes of the fields.  In this theory we must interpret relativistic causality and time ordering in terms of variations (modulations) of four-periodicity rather than of variations of four-momentum (roughly speaking, they are two faces of the same coin).

To see explicitly that the theory is Lorentz invariant we consider a  
global Lorentz transformation   \begin{equation}
x^{\mu}\rightarrow x'^{\mu}=\Lambda_{\nu}^{\mu}~x^{\nu}~
\label{space:Lorentz:tranf}\end{equation}
as a global and isometric substitution of variables in the action (\ref{free:act}). The resulting action  
 \begin{equation}
\mathcal{S}=\oint_{0}^{T'^{\mu}}d^{4}x'\mathcal{L}(\partial'_{\mu}\Phi'(x'),\Phi'(x'))\,\label{Lorentz:trans:action}\end{equation}
describes the same elementary system of (\ref{free:act}) but in a new reference frame. It is important to note that, even though we are considering an isometry and the form of the bosonic Lagrangian  is unchanged,  the transformation of variables induces a correspondent transformation (rotation) of the boundary of the theory
\begin{equation}
T^{\mu}\rightarrow T'^{\mu}=\Lambda_{\nu}^{\mu}~T^{\nu}\,.\label{period:Lorentz:tranf}\end{equation}
Indeed we find that  $T^{\mu}$ ---  being an ordinary space-time interval ---  transforms as a contravariant four-vector. 

 This implies that the field solution $\Phi'(x')$ minimizing the transformed action (\ref{Lorentz:trans:action}) is in general different from the field solution $\Phi(x)$ of the  original action (\ref{free:act}). They have the same equations of motion but they have different BCs. This aspect will play a central role in the rest of the paper.  Besides the naive substitution of the 4D variable $x \rightarrow x'$, to minimize the transformed action (\ref{Lorentz:trans:action}), the transformed field solution $\Phi'(x')$ must have transformed de Broglie four-periodicity $T'^{\mu}$.
Thus, according to (\ref{eq:PdB:relation}), the transformed four-momentum  associated with the transformed field $\Phi'(x')$ is
 \begin{equation}
\bar{p}_{\mu}\rightarrow\bar{p}'_{\mu}=\Lambda_{\mu}^{\nu}~\bar{p}_{\nu}\,.\label{mom:Lorentz:tranf}\end{equation}
This actually describes the four-momentum of our elementary particle in the new reference system, as expected from the Lorentz transformation (\ref{space:Lorentz:tranf}). 
The de Broglie phase harmony (\ref{eq:PdB:relation}) prescribes that $T^{\mu}$ is such that the phase of the field is
invariant under four-periodic translations. But the phase of the field is also invariant (scalar) under transformation of variables. Thus $$ e^{-\frac{i}{\hbar}x^{\mu}\bar{p}_{\mu}}=e^{-\frac{i}{\hbar}(x^{\mu}+cT^{\mu})\bar{p}_{\mu}} \rightarrow e^{-\frac{i}{\hbar}(x'^{\mu}+cT'^{\mu})\bar{p}'_{\mu}}= e^{-\frac{i}{\hbar}x'^{\mu}\bar{p}'_{\mu}} \,.$$  

 The intrinsic four-periodicity of the field  transforms from reference frame to reference frame in a relativistic way as in (\ref{period:Lorentz:tranf}). This description is equivalent to the relativistic Doppler effect.  By using the useful notation $\bar p_{\mu} = h / T^{\mu} c$  \cite{Kenyon:1990fx} the relativistic constrain on the variations of the temporal and spatial components of the de Broglie periodicity can be expressed as \beq
\frac{1}{T_{\tau}^{2}}\equiv\frac{1}{T_{\mu}}\frac{1}{T^{\mu}}\,.\label{rel:constr:4-T}
\eeq
This constrain is induced by  the underlying Minkowski metric $c^{2 }d \tau^{2} = d x^\mu d x_{\mu}$. 
In fact, if multiplied by the Planck constant, (\ref{rel:constr:4-T}) is nothing
but the geometrical description in terms of four-periodicity of the relativistic constraint 
\beq
\bar{M}^{2}c^{2}=\bar{p}^{\mu}\bar{p}_{\mu}\,,
\eeq in agreement with (\ref{eq:PdB:relation}). The mass $\bar M$ is described by the proper time periodicity $T_{\tau}$ of the field: $\bar M = h /  T_{\tau} c^{2}$.  
This corresponds to the  time for light to travel across  the Compton wavelength of the elementary system $\lambda_{s} = c T_{\tau}$. The heavier the mass. the faster the periodicity.  A massless elementary system has infinite (or frozen) proper time periodicity.  A hypothetical bosonic particle with the mass of an electron has  proper time periodicity of the order of $\sim 10^{-20} s$. This is extremely fast even if compared with the modern resolution in time which is of the order of $\sim 10^{-17} s$, or with the characteristic periodicity of the Cesium atom which by definition is $\sim 10^{-10} s$.

The geometric constraint (\ref{rel:constr:4-T}) describes  the relativistic dispersion relation of the energy   of our elementary system 
\beq
\bar{E}(\mathbf{\bar{p}})=\sqrt{\bar{\mathbf{p}}^{2}c^{2}+\bar{M}^{2}c^{4}}\,.\label{fund:mode:disp:relat}
\eeq

In the first part of the paper we will consider  only the fundamental field solution $\bar \Phi(x)$  of the action in compact 4D (\ref{free:act}). This is the  single mode of de Broglie four-angular-frequency $\bar \omega_{\mu}$ associated with the PBCs at the boundary $T^{\mu}$.  We use  the normalization of a string vibrating in compact 4D (similar to the case of ``a particle in a box''), so that  $\bar \mathcal{N}$ depends only on the volume of the compact 4D and  is invariant under isometries.  We will  denote by the bar sign the quantities related to the fundamental mode of a cyclic field.  
    That is,
\begin{eqnarray}
\bar \Phi(x) = 
 \bar \mathcal{N} \bar \phi(x) =  \bar \mathcal{N}  e^{-\frac{i}{\hbar} \bar p_{\mu}  x^{\mu}}~. \label{field:fun:mode} \end{eqnarray}
    As we will see   more explicitly in sec.(\ref{Quantum:Formalism}), this fundamental mode corresponds to the non-quantum (tree-level) limit of the theory. Since its dispersion relation is (\ref{fund:mode:disp:relat}), it describes the relativistic behavior of a corresponding classical particle with mass $\bar M$. 
 In fact, the fundamental solution $\bar{\Phi}(x)$, extended  to the whole Minkowskian space-time $\mathbb{R}^4$, formally corresponds to a single 
mode of an ordinary non-quantized free Klein-Gordon (KG) field $\Phi_{KG}(x)=\bar{\Phi}(x)$
with the same  mass $M_{KG}=\bar{M}$ and  energy ${E}_{KG}(\bar{\mathbf{p}})=\bar{E}(\bar{\mathbf{p}})$. In other words, they have the same four-momentum $\bar p_{\mu}$, four periodicity $T^{\mu}$ and dispersion relation (\ref{fund:mode:disp:relat}). 
 Thus, in terms of the invariant mass $\bar{M}$, the fundamental mode $\bar{\Phi}(x)$ is described by the following action in compact 4D 
\begin{equation}
\bar \mathcal{S}=\frac{1}{2}\int^{T^{\mu}}d^{4}x\left[\partial_{\mu}\bar{\Phi}^{*}(x)\partial^{\mu}\bar{\Phi}(x)-\bar M^{2}\bar{\Phi}^{*}(x)\bar{\Phi}(x)\right]\,.\label{action:fundmode:flat:statmass}\end{equation}
This  action is formally the KG action with boundary $T^{\mu}$. In this case the PBCs are replaced by suitable  SBCs in order to select the single fundamental mode with persistent four-periodicity $T^{\mu}$ ---  we have eliminated the circle from the integral symbol in order to distinguish its solution from more general  field solution of the action with PBCs. 
As we will see in sec.(\ref{Generalization:2:FT}), the study of the variations of BCs proposed here for field theory in compact 4D can be extended to the SBCs of ordinary field theory.   

 When in the last part of the paper we will investigate the quantum behavior of the theory, we will need to use the most generic field solution with intrinsic four-periodicity $T^\mu$. We will address this generic solution as cyclic field. Similarly to a vibrating string or a particle in a box it is easy to figure out that such a generic cyclic field solution  (topology $\mathbb S^{1}$) will be a   sum of the eigenmodes associated with a quantized energy-momentum spectrum.

\section{Interaction\label{sec:Towards-Interactions}}

According to the de Broglie-Planck relation (\ref{eq:PdB:relation})
the local and retarded variations four-momentum  occurring during interactions can be equivalently
written as  local and retarded modulations of de Broglie four-periodicity.  
In  GR, modulations of periodicity of the reference clocks are expressed in terms of deformations of the underlying 4D.  Similarly to GR, we will describe the modulation of periodicity of the cyclic fields during interaction  in terms of deformations of the compact 4D of the theory. For these reasons we will generically denote this description with the  term ``geometrodynamics''.  

\subsection{Geometrodynamics}

In relativistic mechanics a generic interaction scheme  can be formalized  in terms of  corresponding variations of the four-momentum $\bar{p}'_{\mu}(x)$ with respect to the non interacting case $\bar{p}_{\mu}$. That is,    
\begin{equation}
\bar{p}_{\mu}\rightarrow\bar{p}'_{\mu}(x)=e_{\;\mu}^{a}(x)\bar{p}_{a}\,.\label{eq:deform:4mom:generic:int}\end{equation}
 With this notation we mean that the persistent four-momentum $\bar{p}_{\mu}$
in any given interaction point $x=X$ deforms into  $\bar{p}'_{\mu}(x)|_{x=X}$ when we switch
on interaction. Hence the interaction scheme (\ref{eq:deform:4mom:generic:int}) is locally  encoded
 by the tetrad $e_{\;\mu}^{a}(x)$.

Similarly to GR where interaction 
is encoded in a corresponding deformation of the underlying 4D, we will formalize interactions in terms of local diffeomorphisms of the compact 4D.  This means that we will generalize the case of global transformation of variable (\ref{space:Lorentz:tranf})  to the case of local transformations. 
We will therefore  use
field theory in curved 4D. 

To describe this  interaction scheme in terms of cyclic fields we must apply the de Broglie-Planck condition (\ref{eq:PdB:relation}) locally.
It must be noted however that when interaction are concerned, the  local four-periodicity of the interacting field which we will denote from now on by $\tau'^{\mu}(x)$, in general does not coincide with the boundary of the theory at $T^{\mu}(x)$. 
 That is, in general 
$\tau'^{\mu}(x) \neq T^{\mu}(x)$.
As we will see, $T^{\mu}(x)$  transforms as a finite and contravariant four-vector, \emph{i.e.} as $x'^{\mu}$, whereas $\tau'^{\mu}(x)$ transforms as an tangent \cite{Kenyon:1990fx} and contravariant four-vector, \emph{i.e.} as $dx'^{\mu}$. The former describes the recurrence period $\Phi'(x) = \Phi'(x + T)$ whereas the latter describes the \emph{local} or \emph{instantaneous} periodicity in a given point, similarly to the formalism of modulated signals. 
 The local periodicity $\tau'^{\mu}(x)$ of an interacting field varies from point to point, according to the relativistic retarded potentials. In this case  the local de Broglie phase harmony is 
\begin{equation}
{c } \bar{p}'_{\mu}(x) \tau'^{\mu}(x)={h}\,.\label{planck:deBroglie:rel:local}\end{equation}
Therefore the local variation of four-momentum in the interaction scheme (\ref{eq:deform:4mom:generic:int})
 corresponds to the contravariant local variation of the four-periodicity \begin{equation}
T^{\mu}\rightarrow\tau'^{\mu}(x)=T^{a}e_{a}^{\;\mu}(x)\,.\label{eq:deform:4period:generic:int}\end{equation}
Similarly to the variation of four-momentum induced by a given interaction scheme with respect to the free case, the 
persistent four-periodicity $T^{\mu}=\tau^\mu$ of the free field $\Phi(x)$  turns out to be modulated to the local four-periodicity $\tau'^{\mu}(x)$ for the interacting field $\Phi'(x)$.

The deformation of the local four-periodicity $\tau'^{\mu}(x)$ of the  cyclic fields is associated with  the corresponding stretching of the compactification four-vector $T'^{\mu}(x)$ of the theory through the PBCs.
This actually induces  
a deformation of the original Minkowskian  metric according to the following relation
 \begin{equation}
\eta_{\mu\nu}\rightarrow g_{\mu\nu}(x)=e_{\;\mu}^{a}(x)e_{\nu}^{\; b}(x)\eta_{ab}\,.\label{eq:deform:metric:generic:int}
\end{equation}

This resulting  curved space-time encodes the interaction scheme (\ref{eq:deform:4mom:generic:int}) locally. 
To check this geometrodynamical description  we consider a local transformation of variables   \begin{equation}
x^{\mu}\rightarrow x'^{\mu}(x)=x^{a}\Lambda_{a}^{\;\mu}(x)\,, \label{eq:deform:variables:generic:int}\end{equation}
  whose  tetrad matches the one used in (\ref{planck:deBroglie:rel:local}) to encode our interaction scheme. That is 
\begin{equation}
e_{\;\mu}^{a}(X)=\left(\frac{\partial x^{a}}{\partial x'^{\;\mu}}\right)_{x'=X}\,.\end{equation}
 For the scope of this paper we will work in the approximation $e_{\;\mu}^{a}(x')\simeq e_{\;\mu}^{a}(x)$ (for the sake of simplicity we neglect Christoffel symbols). 
Since $x^{a} \Lambda_{a}^{\;\mu}(x)$ is the primitive
of the tetrad $e^{\;\mu}_{a}(x)$ we can use the following notation (omitting
prime indexes in the integrands) \begin{equation}
x^{a}\Lambda_{a}^{\;\mu}(x)\simeq\int^{x^{a}}dx^{a}e_{a}^{\;\mu}(x) \label{finite:infinites:tetrad}\,.\end{equation}

The transformation (\ref{eq:deform:variables:generic:int}) relates locally the inertial frame $x \in  S$ of the free cyclic field solution $\Phi$   to the generic frame $x' \in  S'$ associated with the interacting cyclic field solution $\Phi'$. Its Jacobian is  $\sqrt{-g(x)}=\det[e_{\;\mu}^{a}(x)]$. The Latin letters describe the free field 
 in an inertial frame $ S$ while
the Greek letters refer to the 
 locally accelerated frame $ S'$ of the interacting field $\Phi'$ \cite{Birrell:1982ix}. 
Finally, by using   (\ref{eq:deform:variables:generic:int})    as a substitution of variables in 
 the compact 4D action  (\ref{free:act}), we find that the interaction scheme (\ref{eq:deform:4mom:generic:int}) is
 described by the following action  in locally deformed compact 4D 
\begin{equation}
\mathcal{S}\simeq\oint^{T^{a}\Lambda_{a}^{\mu}|_{X}(T)}d^{4}x\sqrt{-g(x)}\mathcal{L}(e_{a}^{\;\mu}(x)\partial_{\mu}\Phi'(x),\Phi'(x))\,.\label{eq:defom:action:generic:int}\end{equation}
It is important to point out that (\ref{eq:deform:variables:generic:int}) induces 
the local deformation (or stretching) of the compactification four-vector
\begin{equation}
T'^{\mu}(X)\simeq T^{a}\Lambda_{a}^{\;\mu}|_{X}(T) \simeq  \int^{X^{a}+T^{a}}_{X^{a}} dx^{a} e_{a}^{\;\mu}(x)\,.\label{eq:deform:4T:generic:int}\end{equation}
This is  the  local deformation of the boundary associated with the modulation of local periodicity $\tau'^{\mu}(x)$ (\ref{eq:deform:4period:generic:int}), \emph{i.e.} with the interaction scheme (\ref{eq:deform:4mom:generic:int}).

As we well see later the resulting formalism will be the four-dimensional analogous of the formalism of modulated signals. 
Indeed our interacting system  is described by the cyclic field solution $\Phi'(x)$ of the transformed action (\ref{eq:defom:action:generic:int}) in curved space-time (\ref{eq:deform:metric:generic:int}) whose four-frequency or four-momentum spectrum is modulated point by point. 
According to the diffeomorphism (\ref{eq:deform:variables:generic:int}) or to the phase harmony (\ref{planck:deBroglie:rel:local}),
if the free cyclic field $\Phi(x)$ has four-momentum $p_{\mu}$, the  transformed cyclic field $\Phi'(x)$ has corresponding modulated four-momentum
  $\bar{p}'_{\mu}(x)$.
The four-momentum of the field is in fact described by the derivative operator
$\partial_{\mu}$ which transforms as the tangent four-vector $\bar{p}'_{\mu}(x)$, \emph{i.e.} $\partial_{\mu} \rightarrow \partial'_{\mu}=e_{\;\mu}^{a}(x)\partial_{a}$. 

The explicit form of the cyclic field solution $ \Phi'(x)$ will be written later for the  particular transformations  where the normalization  $\bar \mathcal{N}$ is left invariant.  
To study the tree-level behavior of the system under the generic interaction scheme (\ref{eq:deform:4mom:generic:int}) in this first part of the paper will be sufficient to consider only the fundamental modes
$\bar{\Phi}'(x)$.  Similarly to (\ref{action:fundmode:flat:statmass}),  its interaction is described by the transformed action  
\begin{equation}
\bar \mathcal{S}\simeq \int^{T^{a}\Lambda_{a}^{\mu}|_{X}(T)}d^{4}x\sqrt{-g(x)}\bar \mathcal{L}(e_{a}^{\;\mu}(x)\partial_{\mu}\bar \Phi'(x),\bar \Phi'(x))\,.\label{action:fundmode:curved:gen}\end{equation}
This  is formally the KG action in curved space-time $g_{\mu\nu}(x)$ with finite integration region $T'^{\mu}(X)$ and suitable SBCs to select $\bar \Phi'(x)$ as solution.  

\subsubsection{Generalization to ordinary field theory}\label{Generalization:2:FT}

In the practical applications of ordinary field theory the role of the BCs is marginal. The ordinary fields used for computations are the more generic solution of the KG equation, \emph{i.e.}  an integral over all the possible energy eigenmodes. In this paper we will see that the formalism of fields in compact 4D has  interesting  properties that are not manifest in ordinary field theory. The variation of the boundary of the action describes different  field solutions, and in turns different kinematic configurations of the same particles. The action (\ref{action:fundmode:curved:gen}) actually allows one to investigate the dynamical behavior of a single KG mode under transformations of reference frame. Its variation of four-momentum $\bar{p}'_{\mu}(x)$, \emph{i.e.} its kinematics, is therefore encoded in the deformation of the boundary, in a manner of a  holographic principle. 

Although in ordinary field theory BCs are not explicitly used in practical computations, a boundary $\Sigma^{\mu}$ is implicitly assumed (typically of infinite spatial lengths). Suitable SBCs at $\Sigma^{\mu}$ can also be applied  to ordinary KG action  in order to select a particular single mode which locally matches a KG mode: $\bar \Phi'(x)|_{x=X}\equiv \bar \Phi_{KG}(X)$.   The generalization to the ordinary field theory with generic boundary $\Sigma^{\mu}$ and suitable SBCs is formally obtained from  (\ref{free:act}) through the following formal substitution $\oint^{T^{\mu}} \rightarrow \int^{\Sigma^{\mu}}$ \footnote{Typically $\Sigma ^{\mu }$ has infinite size along the spatial directions, at
  whose boundary the field is imposed to vanish, whereas along the time
  dimension the boundaries are the generic initial and final time where SBCs
  are supposed. For a KG theory, given the mass as a parameter,  the description of the deformation of the time boundary with SBCs is sufficient to described
  kinematical variations of the field or interactions.}. 
By analogy to field theory in compact 4D, through the transformation of variable (\ref{eq:deform:variables:generic:int})
it is easy to find that --- in the approximation used in this paper --- this generic integration region $\Sigma^{\mu}(x)$ transforms as (\ref{eq:deform:4T:generic:int}). That is, 
$\Sigma^{\mu}\rightarrow\Sigma'^{\mu}(X) \sim \Sigma^{a} \Lambda_{a}^{\;\mu}|_{X}(\Sigma)$. 
With this we wanted to   point out that the analysis of  gauge interactions that we will carry on in the next sections for the fundamental mode, can be extended to  single modes of ordinary non-quantized KG fields. 

\subsection{Applications}

We have introduced interactions for cyclic fields in terms of the local
diffeomorphisms  (\ref{eq:deform:variables:generic:int}) from a flat manifold $S$ to a generic manifold $S'$. Thus, under the interaction scheme (\ref{eq:deform:4mom:generic:int}), our system   is described in terms of the geometry of the compact 4D. 

In order to illustrate this in a heuristic way, we will briefly discuss two examples: the weak Newtonian interaction and the QGP logarithmic freeze-out. The complete analysis of the former case will be left to forthcoming papers whereas the case of the QGP has been investigated in \cite{Dolce:AdSCFT}.  
In this preliminary section we will work in the approximation in which  the local four-periodicity can be identified with the compactification four-vector of the theory 
$\tau^{\mu}(x) \sim T^{\mu}(x)$. 
 This approximation is  valid in the cases  where the metric varies sufficiently  smoothly, \emph{i.e.} $e_{\;\mu}^{a}(x) \sim \Lambda_{\;\mu}^{a}(x) $ (or logarithmically, see \cite{Dolce:AdSCFT}).

In the next section, by writing (\ref{eq:deform:4mom:generic:int})
as a minimal substitution,  we will show that such a geometric approach to interactions
can be also used to describe gauge interactions. 

\subsubsection{Weak Newtonian potential}\label{weak:Newt:pot}

The geometric approach to interactions described above is interesting because
it actually mimics  the usual geometrodynamical approach
of linearized gravity. To show this we consider a weak Newtonian potential $V(\mathbf{x})=-{GM_{\odot}}/{|\mathbf{x}|}\ll c^2$.
Under this interaction scheme the energy in a gravitational well varies with respect
to the free case as $\bar{E}\rightarrow\bar{E}'\sim\left(1+{GM_{\odot}}/{|\mathbf{x}|c^2}\right)\bar{E}$, \cite{Ohanian:1995uu}.
According to the geometrodynamical description of interaction (\ref{eq:deform:4period:generic:int}),
this means that the de Broglie clocks in a gravitational well (the periodicity of  cyclic fields) are
slower with respect to the free clocks $T_{t}\rightarrow \tau_{t}'\sim\left(1-{GM_{\odot}}/{|\mathbf{x}|c^2}\right)T_{t}$.
With the schematization of interactions based on the de Broglie phase harmony (\ref{eq:PdB:relation}), we have retrieved two important
predictions of GR such as time dilatation and gravitational red-shift $\bar{\omega}\rightarrow\bar{\omega}'\sim\left(1+{GM_{\odot}}/{|\mathbf{x}|c^2}\right)\bar{\omega}$.

Besides this we may also consider the analogous variation of spatial
momentum and the corresponding variation of spatial periodicities of cyclic fields in a gravitational well
\cite{Ohanian:1995uu}. Thus, according to (\ref{eq:deform:metric:generic:int}),
we find that the weak Newtonian interaction turns out to be encoded in the linearized
Schwarzschild metric. 

Indeed, the geometrodynamical description of interaction in the formalism of compact 4D 
actually mimics linearized gravity. In the formalism of compact 4D with PBCs, \emph{i.e.} under the assumption of intrinsic periodicity, every cyclic field can be regarded as a relativistic reference clock (see again Einstein's definition \cite {Einstein:1910}), namely the ``de Broglie internal clock''. The diffeomorphism (\ref{eq:deform:variables:generic:int}) encodes the modulation of periodicity of this clocks occurring  during interaction. This is similar to GR where gravitational interaction can be interpreted in terms of modulations of periodicity of reference clocks encoded in a corresponding deformed 4D background. 
Furthermore, it is well known that GR can be derived from the linearized formulation by considering self-interactions \cite{Ohanian:1995uu} --- for instance by relaxing the assumption of smooth interactions. 
 Nevertheless it is important to mention that   ``what is fixed at the boundary of the action principle of GR''  is not uniquely defined \cite{springerlink:10.1007/BF01889475}. More considerations about this aspect has been given in \cite{Dolce:Dice,Dolce:FQXi}.  SR and GR fix the differential framework of the 4D without giving any particular prescription about the BCs. The only requirement for the BCs is to minimize a relativistic action at the boundary. For this aspect both SBCs and PBCs have the same formal validity and consistence with relativity. With this analysis we have provided an evidence 
 of the consistence of our  formalism of  compact 4D
with GR.  

\subsubsection{AdS/CFT interpretation}\label{AdSCFT}

In \cite{Dolce:AdSCFT}  we have applied the geometrodynamical description of interaction in compact 4D to a simple Bjorken Hydrodynamical Model for QGP logarithmic freeze-out \cite{Magas:2003yp}. In a first approximation the fields constituting the QGP can be supposed massless and
their four-momentum  can be supposed to decay exponentially during the freeze-out. According to (\ref{eq:deform:4mom:generic:int}),  this interaction
scheme is therefore encoded by the conformal warped tetrad $e_{\mu}^{a}(|x|)=\delta_{\mu}^{a}e^{-k |x|/c}$,
where $|x|/c= s/ c = \tau$ and $k$ are the proper time and the gradient of the  QGP freeze-out, respectively.  The time periodicity $T_{t}(s) = e^{ks}/k = h / E(s)$ is the  conformal parameter which describes naturally the inverse of the energy of the fields during the freeze-out, according to de Broglie. 
The resulting variations  of normalization  of the cyclic field solutions  during the freeze-out reproduce formally the logarithmic running of the coupling constant typical of QCD \cite{Dolce:AdSCFT}.
 In this paper we will not explore further the running associated with gauge interactions. This means that we will limit our investigation to  transformations of variables which (in a first approximation) preserve the lengths.  
 
  Moreover, in \cite{Dolce:AdSCFT} we have also shown that a field in deformed compact 4D is dual to a massless XD field in a corresponding 5D metric. For instance they have the same topology $\mathbb{S}^1$.  The dualism is explicit  if we assume that our cyclic world-line parameter plays the role of a cyclic XD. To address this identification we say that the theory has a \emph{virtual} XD, see also \cite{Dolce:2009ce}.
In the \emph{virtual} XD formalism,  the QGP exponential freeze-out turns out to be encoded in a \emph{virtual} AdS metric
and  the classical configurations of cyclic fields in such a deformed
background reproduces basic aspects of AdS/QCD.

\section{Rotating the boundary}\label{Isometry}

In order to give a mental picture of the description of gauge interaction that we want to achieve in next sections we can imagine a charged particle (for simplicity a boson) interacting electromagnetically. Typically, the particle will be characterized by dynamics (similarly to the trembling motion of the \emph{zitterbewegung}) induced by the interaction. Intuitively we will describe such dynamics in terms of local transformations of flat reference frame (avoiding the use of creation and annihilation operators). \footnote{More exactly ``it is natural to attribute the origin of the \emph{zitterbewegung} to the self-interaction of the electron with its own electromagnetic field.'' \cite{Hestenes}} The resulting modulations of de Broglie space-time periodicity will be used to reproduce gauge interaction. The formalism that we will adopt has analogies with that modulated signals. In analogy with an antenna, the EM field can characterized by the  dynamics of the charged particle generating it. 

Indeed, in the particular approximation of local isometries, only the boundary of the theory is deformed  without affecting the underlying flat metric.   
 The coefficients of these local isometries can be described in terms of vectorial fields which therefore encode the transformation. The resulting variation of four-momentum and modulation four-periodicity of the field solution  turns out to be written formally as the  \emph{minimal substitution} and the \emph{parallel transport} of ordinary electrodynamics, respectively.

\subsection{Minimal substitution}

We now want to apply the geometrodynamical description of  interaction  (\ref{eq:deform:4mom:generic:int}) to the following local  infinitesimal transformation of variables 
\begin{equation}
x^{\mu}(x)\rightarrow x'^{\mu}\sim x^{\mu} - e x^{a} \Omega^{\;\mu}_{a}(x)  \,. \label{transf:expans:loc:gener}
\end{equation}
 The coefficient of the expansion is denoted by $e$ and address as \emph{gauge coupling}. In this paper we will work in the approximation in which this transformation is a local isometries, \emph{i.e.}  we  limit our study  to the case  in which the Jacobian is  $\sqrt{-g'} \simeq 1$. 
 
In terms of the formalism  (\ref{eq:deform:variables:generic:int}), a local isometry is described by  
\beq
\Lambda_{\;\mu}^{a} (x) \simeq  \delta_{\mu}^{a} - e \Omega_{\;\mu}^{a}(x)\,,\label{transf:expans:loc:finite}
\eeq
whereas the tangent transformation  \cite{Kenyon:1990fx}  is described by the local tetrad 
\beq
e_{\;\mu}^{a} (x) \simeq  \delta_{\mu}^{a} - e \omega_{\;\mu}^{a}(x)\,. \label{transf:expans:loc:tetrad}
\eeq
As we will discuss in sec.(\ref{Non-abelian-case}),  these isometries can be regarded as belonging to some local subgroups of the Lorentz transformations.  Moreover the requirement of local isometries implies that $\omega_{\;\mu}^{a}(x)$ is antisymmetric (Killing equations). For the sake of simplicity we assume that such an isometry is a unitary transformation $$\omega_{\;\mu}^{a}(x) \in U(1)\,.$$
According to (\ref{finite:infinites:tetrad}) the finite  and tangent transformations are  related by 
\begin{equation}
x^a \Omega^{\;\mu}_{a}(x) = \int^{x^a} d x^a \omega^{\;\mu}_{a}(x)\,.
\end{equation}

To each point $x= X$ of the inertial frame $S$ we are associating a local Lorentz reference frame $S'$, as represented  by the orthogonal tetrad  $e_{\;\mu}^{a} (X)$. 
As already noticed, the tetrad $e_{\;\mu}^{a} (x)$ encodes the information, point by point, of a corresponding  interaction scheme. 
In this case the information is contained in $\omega^{\;\mu}_{a}(x)$ which in turn can be  
used to define  a \emph{vectorial} field $\bar{A}_{\mu}(x)$  as
\begin{eqnarray}
 \bar{A}_{\mu}(x) \equiv \omega_{\;\mu}^{a}(x)\bar{p}_{a}  \,.\label{vectorial:field:definition}
\end{eqnarray}
Thus the vectorial fields $\bar{A}_{\mu}(x)$ can be used to encode the interaction scheme  (\ref{transf:expans:loc:gener}). In particular, the variation of four-momentum (\ref{eq:deform:4mom:generic:int})  
 associated with this local isometry  is in this case given by \begin{eqnarray}
\bar{p}'_{\mu}(x)  \sim  \bar{p}_{\mu}- e\bar{A}_{\mu}(x)\,,\label{minim:subst:gener}
\end{eqnarray}
which formally is the  \emph{minimal substitution} of the vectorial field $\bar{A}_{\mu}(x)$.

Since we are assuming  that  the local isometry (\ref{transf:expans:loc:gener})  is a unitary transformation $\omega(x) \in U(1)$, we say that  the vectorial field $A_{\mu}(x)$ is an \emph{abelian} field. In sec.(\ref{Non-abelian-case}) we will discuss  the geometrical meaning of this assumption as well as  the generalization to \emph{non-abelian} isometries. 

Under such a local isometric change of flat manifold $g_{\mu\nu}(x)\simeq \eta_{\mu\nu}$,  the resulting transformed action (\ref{action:fundmode:curved:gen}) has the same structure of the original one, \emph{i.e.} the equations of motion remain unchanged. Nevertheless the boundary of the transformed action   turns out to be locally rotated with respect to the original action (\ref{action:fundmode:flat:statmass}). In fact, the resulting action is 
\begin{equation}
\bar \mathcal{S}=\frac{1}{2}\int^{T'^{\mu}(X)}d^{4}x\left[\partial'_{\mu}\bar{\Phi}'^{\dagger}(x)\partial'^{\mu}\bar{\Phi}'(x)-\bar M^{2} \bar{\Phi}'^{\dagger}(x)\bar{\Phi}'(x)\right]\,.\label{target:action:flat:fundam}\end{equation}
This action  has indeed transformed boundary at $T'^{\mu}(X)$ with respect to the persistent boundary at $T^{\mu}$ of the original action, according to (\ref{eq:deform:4T:generic:int}).  Therefore  
its fundamental solution $\bar\Phi'(x)$ in the point $x=X$ turns out to be different from the original fundamental  solution $\bar\Phi(x)$.  The free solution $\bar\Phi(x)$ has the persistent periodicity $\tau^{\mu}=T^{\mu}$ of an isolated particle.  The transformed solution $\bar\Phi'(x)$ has modulated periodicity $\tau'^{\mu}(X)$ varying from point to point in order to describe interaction.

\subsubsection{Global Isometry}\label{Glob.Isometry}

To formulate interactions in terms of gauge fields we need  to  express the rotations of the boundary  in terms of \emph{internal transformations} of the field. 

Here, as in par.(\ref{Compact4D:formalism}), we consider the simple case of a \emph{global}  isometry  
\begin{equation}
\Lambda_{\;\mu}^a(x) = e_{\;\mu}^a(x) \equiv e_{\;\mu}^a\,. \label{intern:sym:tetr}
\end{equation} 
For reasons that will be clear later  we address this case as \emph{pure gauge} --- this terminology is not completely equivalent to ordinary QFT.  
 
In this case the tetrad is homogeneous, it does not depend on $x$. The resulting four-periodicity $\tau'^{\mu}=e^{\mu}_{a}T^{a}$ and  four-momentum $\bar p'_\mu=e_{\mu}^{a}\bar{p}_{a}$ of the transformed field $\bar \Phi'$  vary globally. The transformed four-periodicity (\ref{period:Lorentz:tranf}) coincides with the  compactification four-vector of the transformed action $\tau'^\mu=T'^\mu$.

  In every interaction point $x=X$ the fundamental solution $\bar \Phi'(x')$  associated with the transformed action (\ref{target:action:flat:fundam}) is related to the original solution $\bar \Phi(x)$ by the following transformation 
\beq
\bar{\Phi}(x)=\bar \mathcal{N} e^{-\frac{i}{\hbar} x^{\mu}\bar{p}_{\mu}}\rightarrow\bar{\Phi}'(x') =\bar \mathcal{N} e^{-\frac{i}{\hbar} x{'}^{\mu}\bar{p}'_{\mu}}\,.\label{intern:sym:fund}
\eeq
It is easy to see that, as long as $\bar \Phi(x)$ is a fundamental solution of the free action, under this transformation $\bar \Phi'(x')$ is automatically the correct fundamental solution of the transformed action (\ref{target:action:flat:fundam}). In fact, it has transformed four-periodicity  $T'^{\mu}=e^{\mu}_{a}T^{a}$, according to the de Broglie phase harmony.

Now we expand the global transformation $e_{\mu}^a$ as  in (\ref{transf:expans:loc:gener}), so that  the interaction scheme is formally  the   \emph{minimal substitution} (\ref{minim:subst:gener});
  $\bar{A}_{\mu}$ and $\bar{p}'_{\mu}$ are homogeneous (constant). The physical effect of this \emph{pure gauge}  is a global transformation of reference frame.
Thus, under this global  isometry  the field transforms as
\beq
\bar{\Phi}(x)=\bar \mathcal{N}  e^{-\frac{i}{\hbar} x_{\mu}\bar{p}^{\mu}}\rightarrow\bar{\Phi}'(x')
=\bar \mathcal{N}  e^{-\frac{i}{\hbar} x{'}_{\mu}(\bar{p}^{\mu}-e\bar{A}^{\mu})} =\bar V(x')\bar \Phi(x{'}) \label{connct:field:trans:U1}\,.\eeq
This also means that to our transformation of variables  there is associated an \emph{internal transformation} of the field described by
\begin{equation}
\bar V(x)=e^{\frac{ie}{\hbar}x_{\mu}\bar{A}^{\mu}}\,.
\end{equation}
We call $\bar A_\mu$ \emph{gauge connection} and $\bar V(x)$ 
 \textit{parallel-transport}. 
Since we are assuming abelian transformation of variables, the resulting \emph{internal transformation} of the fundamental solution (\ref{connct:field:trans:U1}) generates an abelian \emph{parallel-transport}   $\bar V(x) \in U(1)$.

We also introduce the \emph{covariant derivative} of the transformed field $\bar \Phi'$ as 
\begin{equation}
\partial_{\mu}\bar \Phi (x) = \partial_\mu [\bar V^{-1}(x) \bar \Phi'(x)]  = \bar V^{-1}(x) D_{\mu} \bar \Phi'(x)\,. \label{tuned:field:glob}\end{equation}
Thus the covariant derivative associated with a global isometry is 
 \begin{equation}
\partial_{\mu}\rightarrow D_{\mu}=\partial_{\mu}-\frac{ie}{\hbar}\bar{A}_{\mu}\,. \label{tuned:action:glob}
\end{equation}
 It is important to note that, even though the transformed fundamental solution $\bar \Phi'(x)$ has a transformed four-periodicity $T'^{\mu}$, the  terms  $\bar V^{-1}(x) \bar \Phi'(x)$  and $\bar V^{-1}(x) D_{\mu} \bar \Phi'(x) $ have the same persistent  four-periodicity $T^{\mu}$ as  the original fundamental mode $\bar \Phi (x) $ and its derivative $\partial_{\mu}\bar \Phi (x)$, respectively. 
We will address this important aspect by saying that the inverse of the \emph{parallel transport}, together with the \emph{covariant derivative} in derivative terms,  \emph{tunes} the  periodicity  $T'^{\mu}$ of the transformed fundamental mode $\bar \Phi'(x)$ to the periodicity  $T^{\mu}$ of the free fundamental mode $\bar \Phi(x)$. We will next generalize these considerations to local isometries.

\subsection{Local Isometry}

In the  general case of local isometries, the tetrad $e_{\;\mu}^a(x)$ varies from point to point. Thus  we must take into account that, at every point $x=X$, the transformed fundamental mode  $\bar\Phi'(x)$  
  must have local four-periodicity $\tau'^{\mu}(x)|_{x=X}$ in order to be a solution of the transformed action (\ref{target:action:flat:fundam}) with boundary at $T'^{\mu}(x)|_{x=X}$ given by (\ref{eq:deform:4T:generic:int}).  This also means that the transformed fundamental solution $\bar\Phi'(x)$ must have four-momentum $\bar{p}'_{\mu}(x)_{x=X}$ in order to satisfy  (\ref{planck:deBroglie:rel:local}) in $x=X$.
In the approximation $e_{\;\mu}^a(x) \sim e_{\;\mu}^a(x')$ and considering that the normalization factor $\bar \mathcal{N}$ is invariant under isometries, the fundamental mode $\bar\Phi'(x')$ solution of the transformed action (\ref{target:action:flat:fundam}) can be written, according to our notation (\ref{finite:infinites:tetrad}),  as
\beq
\bar{\Phi}(x) = \bar \mathcal{N}  e^{-\frac{i}{\hbar}x \cdot \bar{p}} \rightarrow\bar \Phi'(x')=\bar \mathcal{N}  e^{-\frac{i}{\hbar}\int^{x'^\mu}d x^\mu  \bar p'_\mu(x)} 
\label{transf:field:solution}\,.
\eeq
To  check this, besides the de Broglie phase harmony, the analogy with the CKM formalism and the modulated signals formalism,
we may note that the  derivative operator $i \hbar \partial_{\mu}$ actually gives the correct transformed four-momentum $\bar p'_\mu(x)$  in the new reference frame 
  \beq
i \hbar \partial_\mu \bar \Phi(x) = \bar p_\mu \bar \Phi(x) \rightarrow i \hbar \partial_\mu \bar \Phi'(x') = \bar  p'_\mu(x') \bar \Phi'(x')\,.
\eeq

According to the definition of the local tetrad $e^{a}_{\;\mu}(x)$  in  (\ref{transf:expans:loc:tetrad}),  the fundamental solution $\bar \Phi'(x')$ of (\ref{target:action:flat:fundam}) is obtained from the free fundamental solution  $\bar \Phi(x)$ by the following \emph{internal transformation} of the field
\beq
\bar{\Phi}(x) = \bar \mathcal{N}  e^{-\frac{i}{\hbar}x \cdot \bar{p}} \rightarrow  \bar{\Phi}'(x') 
= \bar \mathcal{N} e^{\frac{i e}{\hbar}  \int^{x'} dx \cdot \bar{A}(x)}  e^{-\frac{i}{\hbar}x'\cdot \bar{p}} =\bar V(x') \bar \Phi(x')\,.  \label{transf:field:solution:gauge}
\eeq

Hence  the \emph{parallel-transport} $\bar V(x)$ describing the \emph{internal transformation} of fundamental solution under this local abelian transformation of variables is formally a Wilson line of the \text{gauge connection}
 \begin{equation}
\bar V(x) = e^{\frac{ie}{\hbar} \int^{x} dx \cdot \bar{A}(x)}\,. \label{Wilsline:gauge:gen}
\end{equation}
This allows one to pass from a fundamental solution with persistent periodicity $\tau^{\mu}$ to a fundamental solution with transformed local periodicity $\tau'^{\mu}(x)$.
In this way, as long as $\bar \Phi(x)$ is solution of the free action, the modulated field $\bar \Phi'(x')$ is  automatically  solution of the transformed action (\ref{target:action:flat:fundam}).  The vectorial field $\bar A_{\mu}(x)$ describes the resulting modulation of periodicity under the local transformation of reference frame. 

The generalization to  local isometry of the  \emph{covariant derivative} (\ref{tuned:field:glob}) of the transformed field $\bar \Phi'(x)$  is 
 \begin{equation}
\partial_{\mu}\rightarrow D_{\mu}=\partial_{\mu}-\frac{ie}{\hbar}\bar{A}_{\mu}(x)\,.\label{tuned:action:loc}\end{equation}
Also in this more general case  we find that the inverse of the \emph{parallel-transport}, together with the \emph{covariant derivative} in derivative terms, \emph{tunes} the locally varying periodicity  $\tau'^{\mu}(x)$   of the transformed fundamental field $\bar \Phi'$ to the persistent periodicity  $\tau^{\mu}= T^{\mu}$ of the free fundamental field $\bar \Phi$.  This \emph{tuning} through  \emph{parallel transport} will be used to allow a field with  locally varying four-periodicity $\tau'^{\mu}(x)$ to fulfill the variational principle in an action with persistent boundary $\tau^{\mu}= T^{\mu}$.

\subsection{Noether Currents}	\label{Noether:currents}

The formalism of field theory in compact 4D allows one to relate, through the variational principle,  transformation of variables $\delta x$ and    \emph{internal transformations} of the field $ \delta \Phi(x)$. 
  This can be easily seen from the role of the \emph{parallel-transport}  (\ref{Wilsline:gauge:gen}).
   The \emph{internal transformation} associated to the  local abelian isometry 
  (\ref{transf:expans:loc:gener}) is indeed 
 \begin{equation}
 \delta\bar  \Phi(x)=\bar{\Phi}'(x)-\bar{\Phi}(x) = i e  \bar  A_\nu (x) \bar \Phi(x) \delta x^\nu\,.\label{intern:transf:fund}
 \end{equation}

\begin{figure}
\begin{center}{
\def\svgwidth{10cm}
\begingroup
  \makeatletter
  \providecommand\color[2][]{%
    \errmessage{(Inkscape) Color is used for the text in Inkscape, but the package 'color.sty' is not loaded}
    \renewcommand\color[2][]{}%
  }
  \providecommand\transparent[1]{%
    \errmessage{(Inkscape) Transparency is used (non-zero) for the text in Inkscape, but the package 'transparent.sty' is not loaded}
    \renewcommand\transparent[1]{}%
  }
  \providecommand\rotatebox[2]{#2}
  \ifx\svgwidth\undefined
    \setlength{\unitlength}{189.97336426pt}
  \else
    \setlength{\unitlength}{\svgwidth}
  \fi
  \global\let\svgwidth\undefined
  \makeatother
  \begin{picture}(1,0.49934227)%
    \put(0,0){\includegraphics[width=\unitlength]{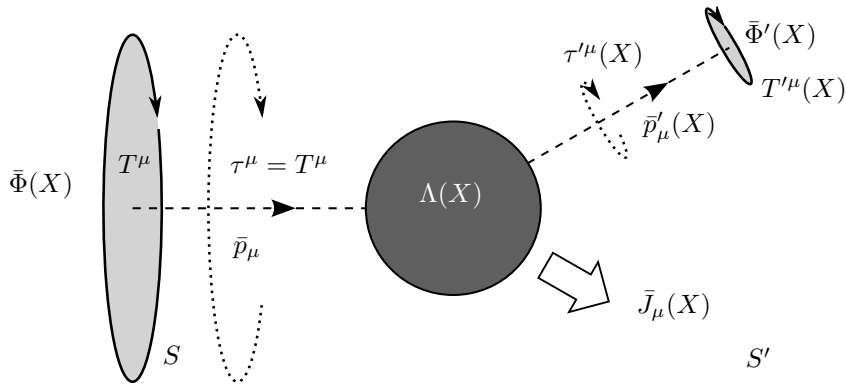}}%
    \put(0.41687792,0.24007029){\color[rgb]{1,1,1}\makebox(0,0)[lb]{\smash{{{$\Lambda(X)$}}}}}%
    \put(0.02103292,0.27797034){\color[rgb]{0,0,0}\makebox(0,0)[lb]{\smash{$T^\mu$}}}%
    \put(0.16842204,0.27797034){\color[rgb]{0,0,0}\makebox(0,0)[lb]{\smash{$\tau^\mu=T^\mu$}}}%
    \put(0.17263316,0.17269242){\color[rgb]{0,0,0}\makebox(0,0)[lb]{\smash{$\bar p_\mu$}}}%
    \put(0.60637811,0.42535942){\color[rgb]{0,0,0}\makebox(0,0)[lb]{\smash{$\tau'^\mu(X)$}}}%
    \put(0.71165603,0.33271486){\color[rgb]{0,0,0}\makebox(0,0)[lb]{\smash{$\bar p'_\mu(X)$}}}%
    \put(0.86746734,0.37903714){\color[rgb]{0,0,0}\makebox(0,0)[lb]{\smash{$T'^\mu(X)$}}}%
    \put(0.70323379,0.09268121){\color[rgb]{0,0,0}\makebox(0,0)[lb]{\smash{$\bar J_\mu(X)$}}}%
    \put(0.8450715,0.45036897){\color[rgb]{0,0,0}\makebox(0,0)[lb]{\smash{$\bar\Phi'(X)$}}}%
    \put(-0.12304447,0.25235537){\color[rgb]{0,0,0}\makebox(0,0)[lb]{\smash{$\bar \Phi(X)$}}}%
    \put(0.08001124,0.02526671){\color[rgb]{0,0,0}\makebox(0,0)[lb]{\smash{{$S$}}}}%
    \put(0.84643451,0.0210556){\color[rgb]{0,0,0}\makebox(0,0)[lb]{\smash{{$S'$}}}}%
  \end{picture}%
\endgroup
 \caption{\label{transf:fig}  
	Diagrammatic description of the local isometry  $\Lambda(x)$ transforming a free fundamental field $\bar \Phi(X)$ of persistent four-momentum $\bar p_\mu$, four-periodicity $\tau^{\mu}$ and compactification four-length $T^{\mu}=\tau^{\mu}$,   from the inertial frame $S$  to the reference frame $S'$. The resulting fundamental field solution $\bar \Phi'(X)$ has  locally transformed four-momentum $\bar p'_\mu(X)$, four-periodicity $\tau'^{\mu}(X)$ and compactification four-length $T'^{\mu}(X)\neq T^{\mu}$. The conservation of stress-energy-momentum tensor 	involves  the Noether current $\bar J_{\mu}(X)$.}}
\end{center}
\end{figure}

  To understand the role of these \emph{internal transformations} of the field we consider the role of the boundary terms in the variation of the action.  In the approximation of local isometries, the original and transformed action differ  by an
 explicit  variation of the boundary
\begin{eqnarray}
\delta \bar \mathcal S = \int^{T'^\mu} d^4 x' \bar \mathcal L' (\bar \Phi'^i,x') -  \int^{T^\mu} d^4 x \bar \mathcal L (\bar \Phi^i,x)\,.\label{Neother:act}
\end{eqnarray}
As represented diagrammatically in fig.(\ref{transf:fig}), the stress-energy-momentum of the fundamental mode $\bar \Phi$
is not manifestly conserved because of the local transformation of reference frame. In fact, it is easy to see from (\ref{Neother:act}) that  the conservation of the stress-energy-momentum tensor involves the contribution of the \emph{internal transformation} of the field (\ref{intern:transf:fund}); that is, of the current 
 \begin{equation}
\bar  J_{\mu} =   ie [\bar \Phi^* D_\mu \bar \Phi -  D_\mu \bar \Phi^*  \bar \Phi]\,.
\label{Noether:current}
 \end{equation}

The interesting aspect of this analysis is that actually the current  $\bar  J_{\mu}$ has the same form as the  Noether current of an abelian gauge invariant theory with  \emph{internal transformation} (\ref{intern:transf:fund}), as long as we identify the connection $\bar A_{\mu}(x)$ with an abelian gauge field.  

\section{Gauge interaction}\label{Gauge:interaction}

The formalism introduced so far is very useful to describe the  geometrodynamics associated with a given interaction scheme of an elementary particle.  However  it does not explicitly take into account the conservation of four-momentum. This is because it involves a transformation of reference frame.  As we have pointed out in the Noether analysis, this is related to the local variations of the boundary. From a analytic  point of view it would be easier to describe the same interaction scheme in terms of an action whose boundary is invariant under isometric transformation of variables. In this way  the  currents related to the transformation of the boundary will be directly expressed in terms of symmetries of the Lagrangian. The possibility of such a formalism is offered by the fact that the periodicities of the fields can be \emph{tuned}  through \emph{parallel transport}. The result will be an ordinary  gauge invariant theory.

\subsection{Tuned action}

We want to define a new formalism to describe  $\bar \Phi'(x)$ under the interaction scheme (\ref{minim:subst:gener}) such that there is an explicit conservation of four-momentum. 
 Our strategy will be  to write  a new action, which we will call \emph{tuned} action, containing the same physical information of the transformed  action (\ref{target:action:flat:fundam})  but with persistent boundary at $T^{\mu}$.   Similarly to the  transformed  action, this \emph{tuned} will be obtained directly from  the free action (\ref{action:fundmode:flat:statmass}). 

 To build the \emph{tuned} action we need to know that  to a change in the periodicity of the fundamental field solution there is  an associated \emph{internal transformation} (\ref{transf:field:solution:gauge}), \emph{i.e} a \emph{parallel-transport}.  
  If we want to vary the four-periodicity of a field  in an action with given boundary, and at the same time  fulfills the variational principle at the boundary,  we must use the \emph{parallel-transport}  to \emph{tune} the four-periodicity of the field.
Since the only terms involved in the BCs are the derivative terms (through integration by parts), the  tuned action can be obtained from the free action by modifying only derivative terms. We have already noticed, for instance, that the corresponding covariant derivative (\ref{tuned:action:loc}) allows one to \emph{tune} the periodicity of $\bar \Phi'(x)$ to the periodicity of $\bar \Phi(x)$.

In the tuned action the interaction will be described in terms of symmetries of  the Lagrangian rather than in terms of the variations of the boundary. Hence the interacting field $\bar \Phi'(x)$ will be described by different equations of motion with respect to the free case and the  currents (\ref{Noether:current}) will be the conserved currents associated with the symmetries of the tuned Lagrangian. 

 \begin{figure}\begin{center}{
\def\svgwidth{10cm}
\begingroup
  \makeatletter
  \providecommand\color[2][]{%
    \errmessage{(Inkscape) Color is used for the text in Inkscape, but the package 'color.sty' is not loaded}
    \renewcommand\color[2][]{}%
  }
  \providecommand\transparent[1]{%
    \errmessage{(Inkscape) Transparency is used (non-zero) for the text in Inkscape, but the package 'transparent.sty' is not loaded}
    \renewcommand\transparent[1]{}%
  }
  \providecommand\rotatebox[2]{#2}
  \ifx\svgwidth\undefined
    \setlength{\unitlength}{195.23679199pt}
  \else
    \setlength{\unitlength}{\svgwidth}
  \fi
  \global\let\svgwidth\undefined
  \makeatother
  \begin{picture}(1,0.48453635)%
    \put(0,0){\includegraphics[width=\unitlength]{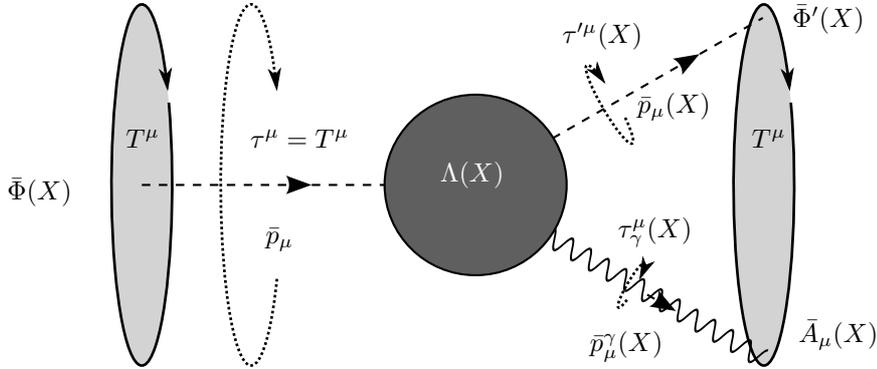}}%
    \put(0.43432071,0.24589353){\color[rgb]{1,1,1}\makebox(0,0)[lb]{\smash{{$\Lambda(X)$}}}}%
    \put(0.02046429,0.290967){\color[rgb]{0,0,0}\makebox(0,0)[lb]{\smash{$T^\mu$}}}%
    \put(0.18436782,0.290967){\color[rgb]{0,0,0}\makebox(0,0)[lb]{\smash{$\tau^\mu = T^\mu$}}}%
    \put(0.59473064,0.4295359){\color[rgb]{0,0,0}\makebox(0,0)[lb]{\smash{$\tau'^\mu(X)$}}}%
    \put(0.84407955,0.290967){\color[rgb]{0,0,0}\makebox(0,0)[lb]{\smash{$T^\mu$}}}%
    \put(0.66788325,0.17213694){\color[rgb]{0,0,0}\makebox(0,0)[lb]{\smash{$\tau_\gamma^\mu(X)$}}}%
    \put(-0.1352204,0.22126974){\color[rgb]{0,0,0}\makebox(0,0)[lb]{\smash{$\bar \Phi(X)$}}}%
    \put(0.89737183,0.45073474){\color[rgb]{0,0,0}\makebox(0,0)[lb]{\smash{$\bar\Phi'(X)$}}}%
    \put(0.905567,0.03278067){\color[rgb]{0,0,0}\makebox(0,0)[lb]{\smash{$\bar A_\mu(X)$}}}%
    \put(0.63102859,0.02048791){\color[rgb]{0,0,0}\makebox(0,0)[lb]{\smash{$\bar p^\gamma_\mu(X)$}}}%
    \put(0.20417481,0.16327876){\color[rgb]{0,0,0}\makebox(0,0)[lb]{\smash{$\bar p_\mu$}}}%
    \put(0.69342627,0.33732941){\color[rgb]{0,0,0}\makebox(0,0)[lb]{\smash{$\bar p_\mu(X)$}}}%
  \end{picture}%
\endgroup
 \caption{\label{tuned:fig}  
 Diagrammatic description of the local isometry  $\Lambda(x)$  in terms of the \emph{tuning} mediated by the vectorial field $A_{\mu}(X)$ with local four-momentum $\bar p^{\gamma}_{\mu}(X)$ and four-periodicity $\tau_{\gamma}^\mu(X)$. The conservation of four-momentum is manifest from the fact that the free fundamental field $\bar \Phi(X)$ and the interacting system  $\bar \Phi'(X)$ and $A_{\mu}(X)$  have the same  compactification length $ T^{\mu}$. 
 }
}\end{center}
\end{figure}

   At a mathematical level this can be easily achieved through the \emph{parallel-transport}  $\bar V(x)$,  by explicitly writing  in the free action (\ref{action:fundmode:flat:statmass}) the  fundamental solution $\bar\Phi(x)$ as a function of the transformed fundamental solution $\bar \Phi'(x)$, \emph{i.e.} by using (\ref{transf:field:solution:gauge}). In this way we find 
 \beq
\int^{T^{\mu}}d^{4}x \bar \mathcal{L}(\partial_\mu \bar \Phi, \bar \Phi ) =  \int^{T^{\mu}}d^{4}x \bar \mathcal{L}(\partial_\mu \bar V^{{-1}}\bar \Phi', \bar V^{{-1}} \bar \Phi' ) = \int^{T^{\mu}}d^{4}x \bar \mathcal{L}(D_\mu \bar \Phi', \bar \Phi' )\,.
 \eeq 
 According to the definition  (\ref{tuned:action:loc}),  the derivative terms $\partial_\mu \bar \Phi(x)$ of the original action must be replaced by the  covariant derivative   $D_\mu  \bar \Phi'(x)$ in order to \emph{tune} locally the  periodicity $\tau'^\mu(x)$ to $T^\mu$.    
We have also used the fact that non-derivative terms, such as  
  the mass term, contributes only to the  equations of motion, but not to the BCs (the periodicity of the solution in such terms  can be arbitrarily varied without compromising the variational principle at the boundary).
     
  Therefore the tuned Lagrangian  can be directly obtained from the free Lagrangian by replacing the ordinary derivatives with covariant derivatives 
 \begin{equation}
\bar \mathcal{L}_{tuned}(\partial_\mu \bar \Phi', \bar \Phi', A_{\mu}) = \bar \mathcal{L}(D_\mu \bar \Phi', \bar \Phi' )\,.
\end{equation}

In the specific  interaction scheme (\ref{minim:subst:gener}) for the fundamental scalar mode of a cyclic field with mass $\bar M$,  the \emph{tuned} action is 
\begin{equation}
\bar \mathcal{S}=\frac{1}{2}\int^{T^{\mu}}d^{4}x\left[D_{\mu}\bar{\Phi}'^{*}(x)D^{\mu}\bar{\Phi}'(x)-\bar M^{2}\bar{\Phi}'^{\dagger}(x)\bar{\Phi}'(x)\right]\,. \label{tuned:act:frat}
\end{equation}
The tuned action is related to the  original free action by \emph{parallel-transport}.  If  $\bar \Phi(x)$ is the fundamental solution of the original free action,  then the transformed fundamental solution  $\bar \Phi'(x)$ automatically minimizes (\ref{tuned:act:frat}).  On the other hand 
 $\bar \Phi'(x')$ is also the solution of the transformed action (\ref{target:action:flat:fundam}).  Hence the tuned action (\ref{tuned:act:frat}) contains the same physical information of the transformed action (\ref{target:action:flat:fundam}).  

 The vectorial field $\bar A_{\mu}(x)$ encodes the modulation of periodicity of an interacting system giving rise to covariant derivatives in the tuned action. 
     If we identify the \emph{gauge connection}  $\bar A_{\mu}(x)$ with  an ordinary gauge field, the tune action (\ref{tuned:act:frat}) is formally a gauged KG action with boundary. 
     
In this way we have given a geometrical meaning to covariant derivatives,  parallel-transports, and gauge connections, in terms of variations of periodicities. This picture can be intuitively interpreted by imagining  the  transporting of the arms of a clock with given periodicity on a closed path in a curved manifold. At the end of the loop we must either \emph{retune} the arms of the clock through \emph{parallel-transport}  or to assume that the clock  has varied its characteristic  periodicity. Indeed the \emph{parallel-transport}  allows one to describe the modulation of periodicity ``associated with deformation of the underlying manifold in a background independent way'', \cite{Ohanian:1995uu}. 

\subsection{Gauge Invariance}

At a mathematical level the \emph{parallel-transport}  $\bar V(x)$ allows one the possibility to \emph{tune} periodicity of terms of the Lagrangian to the periodicity imposed by the boundary of the action. 
However it is easy to note that such a parallel-transport, as well as the gauge connection, is not uniquely defined:  the tuned action (\ref{tuned:act:frat}) has a manifest  invariance which we call --- for obvious reasons --- \emph{gauge invariance}. 

 It is well known that (\ref{tuned:act:frat}) is invariant under the following transformation
\begin{equation}
 \bar A_\mu(x) \rightarrow  \bar A'_\mu(x) = \bar A_\mu(x) - e \partial_\mu \bar \theta(x)\,.\label{gauge:transf:sec}
\end{equation}
In the \emph{parallel-transport}  $\bar V(x)$, which  is formally a Wilson line (\ref{Wilsline:gauge:gen}), this transformation generates boundary terms which can be absorbed by the following local phase transformation of the fundamental scalar mode 
\begin{equation}
\bar \Phi'(x') \rightarrow \bar \Phi''(x)= \bar \mathcal{N} e^{-\frac{i}{\hbar} e \bar \theta(x) }\bar  \Phi'(x)\,.\label{phase:transf:sec}
\end{equation}
The resulting local phase transformation of the field is
\begin{equation}
\bar  U(x) = e^{-\frac{i}{\hbar} e \bar \theta(x) }\,. \label{phase:rot:field:abel}
\end{equation}
This local phase turns out to be of the same kind as the \emph{parallel-transport}  $\bar V(x)$ or of the local isometry generating it. Therefore  $\bar U(x) \in U(1)$ in the case of abelian isometry. We have finally shown that a fundamental mode subject to  a local abelian isometry is described by an abelian gauge invariant action. 

The local gauge invariance $U(1)$ is now a symmetry of the tuned Lagrangian. Therefore gauge transformations do not  affect the boundary of the action.
In other words we have found that gauge invariance identifies particular class of isometries
describing the same interaction scheme. For this reason we call them \emph{gauge orbits} (we only mention that such a gauge orbit can be regarded as an \emph{holonomy}, since it corresponds to an isometry whose parameter $\omega(x)$ is a total derivative and  ``the boundary of a boundary is zero'').  
 
The meaning of the \emph{tuning} of the field described so far can be  generalized. Gauge invariant terms allow one to tune the periodicity of the  fundamental field solution to the periodicity imposed,  through the variational principle, by the boundary of the action. This is essentially because the \emph{parallel-transport}   $\bar V(x)$ tunes the periodicity of the bosonic field  solutions.

  At this point we can repeat the  variational analysis of par.(\ref{Noether:currents}).   Noether's theorem can be applied directly to the  tuned  Lagrangian instead of the transformed action. Since the  tuning formalism allows a description in which  the boundary of the action does not vary under  isometric transformation of variables,  it is easy to show that the Noether currents (\ref{Noether:current}) are directly associated  with the symmetries of the tuned Lagrangian. Indeed the Noether analysis is completely parallel to the one of ordinary gauge theory.  
  
  The remaining step to prove that the interaction scheme associated with such a local abelian isometry is nothing but the usual gauge interaction, is to find that the dynamics of the abelian gauge field $A_\mu(x)$ is actually described by the Maxwell equations.

\subsection{Yang-Mills action}

The tuned action (\ref{tuned:act:frat}) is formally  a $U(1)$ gauge invariant action. The tuning of the interacting field has been obtained at the expense of the  introduction of a new field in the theory. This is the gauge connection $\bar A_{\mu}(x)$. It compensates the variation of four-periodicity of the interacting field in order to have a \emph{tuning} to the persistent boundary of the action. 
 Therefore it is natural to interpret  $\bar A_{\mu}(x)$  as a new dynamical field with given four-momentum, in general different from $\bar p_\mu$ or $\bar p'_\mu(x)$, and thus with  given local four-periodicity, in general  different from $T^\mu$ or $\tau'^\mu(x)$. This is illustrated in fig.(\ref{tuned:fig}).   This requires one to introduce a kinetic term for  $\bar A_{\mu}(x)$.    Similarly to  the interacting fundamental field solution $\bar \Phi'(x)$ of the tuned action (\ref{tuned:act:frat}), this kinetic term may appear in an action with persistent integration region $T^\mu$, in order to describe explicitly the conservation of four-momentum. That is, we want to add a  kinetic term for $\bar A_{\mu}(x)$  to the tuned action (\ref{tuned:act:frat}) --- a similar analysis can be done for the transformed action. 

From the correspondence between gauge invariance and the periodicity tuning,  we require that  such a kinetic term must be  gauge invariant. In fact, only in this way the periodicity of $\bar A_{\mu}(x)$ can  be tuned  with the BCs imposed by  the tuned action. In particular we have already noticed that such a tuning of the periodicity of the field in derivative terms, such as in the kinetic terms, is possible by using covariant derivatives. Through suitable covariant derivatives the four-periodicity of $\bar A_{\mu}(x)$ in the kinetic term can be tuned to $T^{\mu}$.  

Therefore, by following the same requirement of gauge invariance as in ordinary field theory, we infer that  the correct kinetic terms allowed for $\bar A_{\mu}(x)$ by the variational principle is the gauge invariant term $-\frac{1}{4} \bar F_{\mu\nu}\bar F^{\mu\nu}$, where $F^{\mu\nu}$ is the field strength. In fact, it is well known that  the derivatives in the field strength  can be replaced by arbitrary covariant derivatives of the gauge field itself
$
\bar F^{\mu\nu}(x) 
= D'_\mu \bar A_\nu(x) - D'_\nu \bar  A_\mu(x)
$. 
This means that  in such a kinetic term the periodicity of the gauge field is tuned to  the characteristic periodicity of the action through an appropriate covariant derivative (in general different from the one  of the matter field \footnote{For instance, in a given gauge $\bar A_{\mu}(x) $, the correct covariant derivative to tune the gauge field is given by combining the inverse \emph{parallel-transport}  $\bar V^{-1}(x)$ and the  gauge transformation $\bar U(x)$ where $e \bar \theta(x)= \bar p_{\mu} x^\mu$.}).

Finally, consistently with the variational principle, the full action describing the interaction scheme (\ref{transf:expans:loc:gener})  is  
\beq
\bar \mathcal{S}_{YM}=\int^{T^{\mu}}d^{4}x\left\lbrace-\frac{1}{4}\bar F_{\mu\nu}(x) \bar F^{\mu\nu}(x) \right. + \left. \frac{1}{2}\left[D_{\mu}\bar{\Phi}'^{*}(x)D^{\mu}\bar{\Phi}'(x)-\bar M^{2}\bar{\Phi}'^{*}(x)\bar{\Phi}'(x) \right]\right\rbrace\,.\label{YM:action:flat:fundam}
\eeq
We have obtained nothing but the usual non-quantum (tree-level) abelian Yang-Mills action (in a finite volume) describing a matter field solution with four-momentum $\bar p'_{\mu}(x)$ interacting with a gauge field $\bar A_{\mu}(x)$ and total four-momentum $\bar p_{\mu}$.

 It must be noticed that the simultaneous minimization of the above action at the boundary for both the fields $\bar{\Phi}'(x)$ and $\bar A_{\mu}(x)$, or equivalently the requirement of gauge invariance of the action,  constrains  the general form of the gauge field. In turn the so far generic form of the abelian transformation of variables (\ref{transf:expans:loc:tetrad}) is constrained to a particular subclass, modulo gauge orbits. 
   In fact the fundamental field solution $\bar A_{\mu}$   is no more a generic vectorial field.  That is, according to the variational principle, it must be solution of the  equations of motion associated with (\ref{YM:action:flat:fundam}), \emph{i.e.} it has Maxwell dynamics.   
   For instance this means that $\bar A_{\mu}(x)$ is massless and  has only two transversal \emph{d.o.f.}.  On the other hand, through (\ref{vectorial:field:definition}), the Maxwell dynamics of gauge field $\bar A_{\mu}(x)$ corresponds to  related dynamics of the coefficients $\omega_{\;\mu}^a(x)$ of the transformation of variables. Only isometries (\ref{transf:expans:loc:tetrad}) satisfying this dynamics and which we will address as \emph{transversally polarized} are allowed by the variational principle. 
   The result is a geometrodynamical description of ordinary gauge interactions. 

  The kinetic term of the gauge field has been inferred by noticing that  the variational principle requires gauge invariant terms. The requirement of gauge invariance  is actually the usual way to introduce such  kinetic terms in ordinary  YM theory.
  The same argument can be used to introduce the kinematic terms for the gauge field in the formalism of the transformed action (\ref{target:action:flat:fundam}). In this case we must assume  covariant derivatives only in the field strength such that the periodicity of the gauge field is tuned to the locally varying periodicity $\tau'^\mu(x)$ of the interacting field.    
  
``The modern viewpoint'', \cite{Peskin:1995ev,Blagojevic:2002du}, to introduce gauge interaction in ordinary field theory  is to postulate a \emph{parallel-transport}  (\ref{Wilsline:gauge:gen}) --- sometimes called \emph{connection}--- to a matter field, \emph{i.e.} to postulate \emph{internal} symmetries. In this way the derivative terms generate  covariant derivatives and the Lagrangian  is gauge invariant. Thus, in ordinary field theory, ``\emph{the covariant derivative and the transformation law of the connection $\bar A_\mu$ follow from the postulate of local phase rotation symmetry}'' \cite{Peskin:1995ev}. \emph{From the viewpoint of field theory in compact 4D the same gauge invariant Lagrangian is obtained from the geometrodynamics allowed by the variational principle, without postulating it}.  
 In fact we have seen that  the \emph{parallel-transport}  of a matter field arises naturally to describe the modulation of periodicity associated with  local transformation of variables and to tune the periodicity of the field solution to the one imposed by the minimization of the action at the boundary.   The invariance of the action under transformation of variables induces an \emph{internal transformation} of the field solutions.  This reveals an intuitive  geometrodynamical nature of gauge interactions. 
 This important and non-trivial result is in the spirit of Weyl's original proposal of a geometric interpretation of gauge interactions.     
   
The formalism of compact 4D makes manifest this geometrodynamical interpretation of gauge interaction, because it explicitly relates \emph{internal transformations} of the  field  solution to variations of the BCs. The same arguments can be in principle repeated in ordinary field theory by replacing the compact integration region $T^\mu$ with $\Sigma^\mu$ and the PBCs with SBCs. 
  
\subsection{Non-abelian case}\label{Non-abelian-case}

We conclude this section by giving a generalization of our interaction scheme to \emph{non-abelian} local transformations of variables and discussing the relation to space-time symmetries. 

To generalize our result to the non-abelian case we must assume that the transformation of variables (\ref{transf:expans:loc:finite}) originating our interaction scheme (\ref{minim:subst:gener}) is an element of a non-abelian group $H$.
This implies that in the equation obtained so far  we must perform the following substitution from an abelian vectorial field to a non-abelian one
 \begin{eqnarray*} 
\bar A_{\mu}(x) \rightarrow \bar A^{i}_{\mu}(x)\tau^{i}\,,
\end{eqnarray*}
where $\tau^{i}$ are  generators of $H$. 
As a consequence of the commutation relations of the generators the \emph{parallel-transport}  must be written as an path-ordered Wilson line
\begin{equation}
\bar V(x') = e^{\frac{i e}{\hbar} \int^{x'} dx \cdot \bar{A}(x)} \rightarrow \bar V(x') = \mathrm{P} [e^{\frac{i e}{\hbar} \int^{x'} dx \cdot \bar{A}^{a}(x)\tau^{a} }]\,. \label{Wilsline:gauge:abel}
\end{equation}
A similar redefinition must be considered for the covariant derivatives.  In this way (\ref{YM:action:flat:fundam}) turns out to describe  a Yang-Mills theory with non-abelian gauge invariance $H$.

We have described gauge interaction in the approximation where the transformation of variables (\ref{transf:expans:loc:gener}) is a local isometry. In this approximation the lengths of the four vector are preserved and we do not consider variations of the normalizations of the fields \footnote{Weyl's proposal was based on conformal
 transformations. In future papers we will extend our analysis
  to  transformations between two
   flat 4D with different norm, \emph{i.e.} isomorphisms. We expect to find a running of the coupling constant, similarly to what we have already observed for the QGP
  freeze-out \cite {Dolce:AdSCFT}.}. Moreover the variational principle allows  only a particular subclass of  \emph{polarized} local isometries, modulo \emph{gauge orbits}.   In this way the  dynamics for the related vectorial field turns out to be the usual Maxwell's dynamics. 

We can now associate the \emph{polarized} isometries (\ref{transf:expans:loc:finite}) describing our interaction scheme  to corresponding \emph{polarized} local Lorentz transformations. By  representing the Lorentz group as $SU_L(2)\otimes SU_R(2)$,  the most general global isometry which we may consider is  $\omega \in SU_L(2)\otimes SU_R(2)$. In this case $e \omega_{\mu}^{a}= g w_{\; \mu}^{a i} \tau^i + g'  y_{\;\mu}^{a i} \tau^{i}$ where 
 $w_{ \mu}^{a i}$ and $y_{\mu}^{a i}$,  $g$ and $g'$  are the coefficients and the coupling of $SU_L(2)$ and $SU_R(2)$, respectively.  The index $i=1,2,3$ is associated with the generators $\tau^i$ of $SU(2)$. 
 The global components of this isometry describe \emph{pure-gauge}  transformations (global Lorentz transformations) whereas the local polarized components correspond to the gauging of the corresponding subgroup $H \subset SU_L(2)\otimes SU_R(2)$. 
For instance the abelian gauge theory describing ordinary classical electrodynamics can be obtained by assuming that only the corresponding polarized and abelian Lorentz transformations with generator in  $U_{em}(1) \subset SU_L(2)\otimes SU_R(2)$ are local. 
 
Similarly  we can imagine to describe  electroweak interactions by assuming that the local polarized isometries are only those associated with the Lorentz subgroups $SU_L(2)\otimes U_Y(1) \subset SU_L(2)\otimes SU_R(2)$ (the  gauging the electroweak group from a larger global group $SU_L(2)\otimes SU_R(2)$ is typical  in technicolor models and  useful to describe the custodial symmetry of the Standard Model of electroweak interactions, see for instance  \cite{Casalbuoni:2005rs}).  
As for the abelian case (\ref{vectorial:field:definition}) we can define gauge fields associated with $SU_L(2)$ and $U_Y(1)$ as $W_{\mu}(x) = W_{\mu}^{i}(x) \tau^{i} = w_{ \;\mu}^{a i}(x) \tau^i \bar p_a$ and $Y_{\mu}(x) =Y_{\mu}^{3}(x)\tau^{3} = y_{\; \mu}^{a 3 }(x) \tau^3 \bar p_a $. 
Indeed we have the remarkable possibility to relate the electroweak gauge group to a local Lorentz subgroup.

{A possible generalization of this geometrodynamical description of gauge interaction to fermionic fields has a natural realization on the \emph{Zitterbewegung}  models. Originally proposed by Schr\"odinger this idea provides a semi-classical interpretation of the spin and of the magnetic momentum  in terms of cyclic dynamics whose periodicity is actually the de Broglie periodicity of the fermions, see for instance \cite{Sidharth:2008te}. Such a  trembling motion of the fermions can be derived from the Dirac equation.}

Finally, it is interesting to mention that the  geometrodynamical description of gauge interaction described so far has a deep motivation in the so called Kaluza's miracle \cite{Kaluza:1921tu}. This can be seen  by considering the dualism of the theory to an XD theory \cite{Dolce:2009ce,Dolce:AdSCFT}. Under this dualism, the metric (\ref{eq:deform:metric:generic:int}) associated to the substitution of variable (\ref{transf:expans:loc:gener}) turns out to be a Kaluza-like XD metric. This point out interesting algebraic properties of both the curvature tensor and the electromagnetic field tensor as already noticed by Rainich \cite{Rainich} and then improved by Misner and Wheeler \cite{Misner:1957mt}.   

Interesting aspects of the geometrical interpretation of gauge interaction given here, such as \emph{parallel transport}, \emph{holonomy} and the relation with space-time symmetries, have a similar description in the ``Higher Gauge Theory'' of Baez, see for instance \cite{Baez:2010ya}. In an appealing formalism  it is in fact shown how gauge interaction can be described as ``change in phase of a quantum particle''.  

\section{Quantum Behavior}\label{Quantum:Formalism}

So far we have limited our study to the  fundamental mode $\bar \Phi(x)$ of the cyclic field $\Phi(x)$. This  corresponds to study the theory at tree-level. In fact the fundamental mode can be matched with a corresponding single mode of a non-quantized KG field with corresponding four-frequency.    
Because of this matching with  a KG mode, the fundamental mode  can be  in principle quantized in the usual way by imposing explicitly commutation relations and obtaining ordinary scalar QED. 
However, as shown in \cite{Dolce:2009ce} and summarized in this section,  the classical evolution of a  cyclic field (with all its energy excitations) has remarkable correspondences with the quantum evolution of ordinary second quantized fields.  This correspondence has been checked  explicitly in  both the canonical formulation of QM (the theory has implicit commutation relations) and the Feynman Path Integral (FPI) formulation, as well as for many other peculiar quantum phenomena and problems (including Schrodinger problems \cite{Dolce:2009ce,Dolce:2009cev4,Solovev:2010}). The correspondence with the FPI  formulated for free systems in \cite{Dolce:2009ce}, will be generalized to the interacting system studied so far. In fact, as for the free case, an interacting cyclic field also has  Markovian evolution and an Hilbert space can be defined locally. The assumption of PBCs  can be regarded as a semi-classical quantization condition for relativistic fields.

\subsection{Mode expansion}

To investigate the quantum behavior of the theory we  need to use the most generic cyclic field solutions  of the action in compact 4D and PBCs.  By considering the discrete Fourier transform associated with the  cyclic 4D  of the theory,  it is easy to figure out that such a periodic field has a quantized energy-momentum spectrum. For the sake of simplicity we consider the simple topology $\mathbb S^{1}$ so that the spectrum is described by a single quantum number $n$. 
  
From the relation $\bar{E}(\bar{\mathbf{p}})=\hbar\bar{\omega}(\mathbf{\bar{p}})$, the intrinsic time periodicity $T_{t}(\bar{\mathbf{p}})$ of cyclic field in a given reference  frame denoted by $\mathbf{\bar{p}}$ implies 
a quantized energy spectrum $E_{n}(\bar{\mathbf{p}})   = \hbar {\omega_{n}}(\mathbf{\bar{p}})$. In the free case PBCs yield ${\omega_{n}}(\mathbf{\bar{p}})=n\bar{\omega}(\mathbf{\bar{p}})$, which 
is nothing but the harmonic frequency spectrum 
of a vibrating string with the characteristic time periodicity of the field\footnote{The theory can be regarded as a particular kind of string theory. The compact
  world-line parameter plays the role the compact world-sheet parameter of ordinary string theory.
  Therefore it would be more appropriate to speak about strings rather than
  fields.
}. Thus this formulation can be regarded as  the full relativistic analogous to the semi-classical
quantization of a ``particle'' in a box. It also shares deep analogies
with the Matsubara and the Kaluza-Klein (KK) theories \cite{Matsubara:1955ws,Dolce:AdSCFT}.
Indeed, from this harmonic spectrum  and from (\ref{fund:mode:disp:relat}) we see that  free  cyclic fields reproduce  the same quantized energy spectrum of ordinary second quantized
fields (after normal ordering)
\beq
\bar{E}_{n}(\mathbf{\bar{p}})=n\sqrt{\bar{\mathbf{p}}^{2}c^{2}+\bar{M}^{2}c^{4}}\,.\label{energy:spectrum:KG}
\eeq
A free cyclic field $\Phi(x)$ --- here $x^{\mu}=\{c t, \mathbf{x}\}$ --- with intrinsic periodicity $T_{t}(\bar{\mathbf{p}})$, solution of the  action in compact 4D (\ref{Lorentz:trans:action}), 
is a  tower of energy eigenmodes $\phi_{n}(x)$ with eigenvalues (\ref{energy:spectrum:KG}). 
Furthermore, the time periodicity induces a periodicity $\lambda_x$ to the modulo of the spatial dimensions according to (\ref{rel:constr:4-T}). Hence, similarly to the quantization of the energy spectrum, there is a quantization of the modulo of the spatial momentum $|{\mathbf{p}}_n|= n|{\mathbf{p}}| = n h/ \lambda_x$. 

The fact that $T^{\mu}$ is a four-vector means  that there is a single periodicity induced between the time dimension and the modulo of the spatial dimension. In the simple case of topology $\mathbb S^1$, this implies that the resulting quantization is to be expressed in terms of the single quantum number $n$.   It is important to note, however, that together with this  single fundamental periodicity $\mathbb S^1$, in spherical problems (such as the Hydrogen atom, the 3D harmonic oscillator, or problems with particular bounded geometries),  two other cyclic variables (or their deformations) must be considered: the spherical angles $\theta \in [0, 2\pi)$ and $\psi \in [0, \pi)$.  As well-known from ordinary QM they lead to the ordinary quantization of  angular momentum. In this case the cyclic field would be written as a sum over two additional quantum numbers, typically  denoted by $(m,l)$, and the topology of the fields would be $\mathbb S^{1}\otimes \mathbb S^{2}$ (PBCs for the transverse mode of a field, as in the front-light-quantization,  can be used to calculate semi-classically important predictions of perturbative QED such as the anomalous-magnetic-momentum, see for instance \cite{Honkanen:2010nt}). 
 For simplicity we will not consider  the expansion in spherical harmonics (or their deformations). We will investigate only the quantization of the four-momentum spectrum associated with $\mathbb S^{1}$, which in the free case is harmonic  \beq p_{n \mu}= n \bar p_{\mu}\,. \label{quant:4P:free}\eeq 

Depending whether we want to make explicit the normalization factor of the energy eigenmodes, we write the cyclic field solution by using the following notations,  \begin{eqnarray}
\Phi(x)  
= \sum_{n}  \Phi_{n}(x) =  
 \sum_{n}\mathcal{N}_{n}\alpha_{n}(\bar{\mathbf{p}})\phi_{n}(x)  
 = \sum_{n}\mathcal{N}_{n}\alpha_{n}(\bar{\mathbf{p}})e^{-\frac{i}{\hbar} p_{{ n}} \cdot x}~. \label{field:exp:modes} \end{eqnarray}
 As already mentioned we assume that the normalization factor  $\mathcal{N}_{n}$  is invariant under isometries. The coefficients of the Fourier expansion are represented by $\alpha_{n}(\bar{\mathbf{p}})$.    The non-quantum limit corresponds to  the case where  the cyclic field solutions can be approximated with the fundamental mode $\Phi(x) \sim  \bar \Phi(x) = \bar \mathcal{N} \bar \phi(x)$, \emph{i.e.} when only the fundamental level is largely populated: $\bar \alpha(\bar{\mathbf{p}}) \sim 1$ and $\alpha_{n\neq 1}(\bar{\mathbf{p}}) \sim 0$, see \cite{Dolce:2009ce,Dolce:2009cev4}.
 
\subsection{Correspondence with Quantum Mechanics}

Next we summarize the correspondence of field theory in compact 4D with QFT described in \cite{Dolce:2009ce,Dolce:2009cev4}. We note that the evolution of the free cyclic field (\ref{field:exp:modes}) along the compact time dimension is described by the so-called ``bulk''
equations of motion 
$ 
(\partial_{t}^{2}+\omega_{n}^{2})\phi_{n}(\mathrm{x},t)=0
$.  For the sake of simplicity in this section we assume a single spatial
dimension $\mathrm{x}$. 
Thus the time evolution of the energy eigenmodes $\phi_{n}(\mathrm{x},t)$ can be written as
first order differential equations \beq i\hbar\partial_{t}\phi_{n}(\mathrm{x},t)=E_{n}\phi_{n}(\mathrm{x},t)~. \eeq
 The cyclic field (\ref{field:exp:modes}) is a sum of eigenmodes of an harmonic system.
 Actually this  harmonic system is the typical classical system which can be descried in a Hilbert
space. In fact, the energy eigenmodes of a cyclic field form a complete
set with respect to the  inner product \begin{equation}
\left\langle \phi_{n'}(t')|\phi_{n}(t)\right\rangle \equiv\int_{}^{\lambda_{\mathrm{x}}}\frac{{d\mathrm{x}}}{{\lambda_{\mathrm{x}}}}\phi_{n'}^{*}(\mathrm{x},t')\phi_{n}(\mathrm{x},t)\,.\label{inner:prod}\end{equation}    
 They energy eigenmodes can also be used to  define  Hilbert eigenstates
\beq {\left\langle {\mathrm{x},t}|\phi_{n}\right\rangle} \equiv \frac{\phi_{n}(\mathrm{x},t)}{\sqrt\lambda_\mathrm{x}}\,. \eeq

As obvious in the non-interacting case where the cyclic fields have persistent periodicity, 
 the integration region $\lambda_\mathrm{x}$ can be extended to an arbitrary large  integer number of periods $V_\mathrm{x}= N' \lambda_\mathrm{x}$. That is, by assuming $N' \rightarrow \infty$, it can be extended to an infinite region.

In this Hilbert space we can formally build an Hamiltonian operator defined as \beq \mathcal{H}\left|\phi_{n}\right\rangle \equiv\hbar\omega_{n}\left|\phi_{n}\right\rangle\,.\label{Hamiltonian:Oper:def} \eeq
Similar considerations hold for the spatial dimension and the momentum operator is defined as  \beq \mathcal{P}\left|\phi_{n}\right\rangle \equiv-\hbar k_{n}\left|\phi_{n}\right\rangle \,,\eeq
where $k_{n}=n\bar{k}=nh/\lambda_{\mathrm{x}}$. 
An Hilbert state is defined  as generic superposition of energy eigenmodes  
 \beq |\phi\rangle\equiv\sum_{n } a_{n}\left|\phi_{n}\right\rangle \,. \eeq
This means that  cyclic fields can be repressented in an Hilbert space.
With this Hilbert notation  the time evolution of a generic Hilbert state, as well as of a generic cyclic field, turns out to be described by the familiar Schr\"odinger equation \begin{equation}
i\hbar\partial_{t}|\phi(\mathrm{x},t)\rangle=\mathcal{H}|\phi(\mathrm{x},t)\rangle.\end{equation}

As can be easily seen in the free case, \textit{i.e.} homogeneous Hamiltonian (we will generalize later to interactions), the finite time evolution is given by the 
operator \beq \mathcal{U} (t;t_{i})=e^{-\frac{{i}}{\hbar}\mathcal{H}(t-t_{i})}\label{finite:time:evol}\eeq
which turns out to be a Markovian (unitary) operator: \beq \mathcal{U}(t_{f};t_{i})=\prod_{m=0}^{N-1}\mathcal{U}(t_{i}+t_{m+1};t_{i}+t_{m}-\epsilon) \eeq
where $N\epsilon=t''-t'$. Similarly considerations can be applied to  the spatial evolution. 
The finite classical space-time evolution in the Schr\"odinger representation  is 
\beq
|\phi(t,\mathrm{x})\rangle = \mathcal{U}(\mathrm{x},t;0,0) |\phi\rangle = e^{-\frac{i}{\hbar} (\mathcal{H}t -\mathcal{P}\mathrm{x})}|\phi\rangle =  \sum_n a_n e^{-\frac{i}{\hbar} p_n \cdot \mathrm{x} }|\phi_n \rangle \,. \label{Schrod:repr:free} \eeq

In order to allow an easy generalization of the following result to the interaction case, 
we will describe the finite space-time evolutions in terms of the infinitesimal space-time evolutions of the Markovian operator. In particular this will be useful in the interaction case where in every point a different inner product must be considered. As a result we  will obtain the  integral product  $\int_{V_{\mathrm{x}}} \mathcal D \mathrm{x} $.    

All the elements necessary  to build a FPI are already present in the theory without any further 
 assumption than PBCs. In fact, we can plug the completeness relation of the energy eigenmodes in between the elementary time Markovian evolutions obtaining 
\beq
\mathcal{Z} 
=\int_{V_{\mathrm{x}}}\left(\prod_{m=1}^{N-1}{d\mathrm{x}_{m}}\right) U({\mathrm{x}}_{f},t_{f};{\mathrm{x}}_{N-1},t_{N-1})\times \label{int:evol:oper}
  \dots\times  U({\mathrm{x}}_{2},t_{2};{\mathrm{x}}_{1},t_{1})  U({\mathrm{x}}_{1},t_{1};{\mathrm{x}}',t')\,.\label{mark:element:PI}
 \eeq
From the notation (\ref{Schrod:repr:free}),  the elementary space-time
evolutions of a free system can be written as
\beq
 U(\mathrm{x}_{m+1},t_{m+1};\mathrm{x}_{m},t_{m}) 
  =  \left\langle \phi\right|e^{-\frac{i}{\hbar}(\mathcal{H}\Delta t_{m}-\mathcal{P}\Delta \mathrm{x}_{m})}\left|\phi\right\rangle \,,\label{elemet:phase:space}
 \eeq
 where $\Delta \mathrm{x}_{m}=\mathrm{x}_{m+1}-\mathrm{x}_{m}$ and $\Delta t_{m}=t_{m+1}-t_{m}$. Thus, proceeding in a completely standard way we formally obtain the ordinary Feynman Path
Integral (FPI) for the time-independent Hamiltonian  (\ref{Hamiltonian:Oper:def}),
\begin{equation}
\mathcal{Z}=\lim_{N\rightarrow\infty}\int_{V_{\mathrm{x}}}\left(\prod_{m=1}^{N-1}{d\mathrm{x}_{m}}\right)\prod_{m=0}^{N-1}\left\langle \phi\right|e^{-\frac{i}{\hbar}(\mathcal{H}\Delta t_{m}-\mathcal{P}\Delta \mathrm{x}_{m})}\left|\phi\right\rangle\,. \label{periodic:path.integr:Oper:Fey}
\end{equation}
This remarkable result has been obtained by using classical-relativistic mechanics and without any further assumption than intrinsic four-periodicity.
As in the usual FPI formulation in phase-space we are assuming on-shell elementary
space-time evolutions \cite{Feynman:1942us}. 
Though we started with homogeneous $\mathcal{H}$, this derivation, being based on elementary evolutions of a Markovian operator, can  be generalized to the interacting case as we will see in sec.(\ref{FPI-QED}). 

Proceeding in complete analogy with the ordinary derivation of the FPI in configuration space we note  that the infinitesimal products of  (\ref{elemet:phase:space})  in (\ref{periodic:path.integr:Oper:Fey}) can be generically written in terms of the action of the corresponding classical particle 
 \begin{equation}
{\mathcal{S}}_{cl}(t_{f},t_{i})\equiv\int_{t_{i}}^{t_{f}}dt {L}_{cl}  = \int_{t_{i}}^{t_{f}} dt \left(\mathcal{P}\dot \mathrm{x} - \mathcal{H} \right)\,. \label{action:operator:generic}
\end{equation} 
Finally the FPI in (\ref{periodic:path.integr:Oper:Fey})
can be written in the familiar form \begin{equation}
\mathcal{Z}=\int_{V_{\mathrm{x}}} \mathcal{D}\mathrm{x} e^{\frac{i}{\hbar}\mathcal{S}_{cl}(t_{f},t_{i})}\,.\label{eq:Feynman:Path:Integral}\end{equation}

This important result can be intuitively interpreted by considering that  a cyclic field has topology $\mathbb{S}^{1}$. Indeed,  for a field in such a cyclic geometry
there is an infinite set of possible classical paths with different
winding numbers linking every given initial and final configuration. Contrary to  ordinary fields, a cyclic field can self-interfere and this is described by the ordinary FPI \cite{Dolce:2009ce,Dolce:2009cev4}.  

We may also mention that, \cite{Dolce:2009ce}, by evaluating the expectation value of the observable $  \partial_{\mathrm{x}} F(\mathrm{x})$ associated  with the  inner product (\ref{inner:prod}) of our Hilbert space,  integrating by parts, and considering that the boundary term vanishes because of the periodicity of spatial coordinate, we  find
	   \begin{equation}
\left\langle \phi_{f} | \partial_{\mathrm{x}} \mathcal{F}(\mathrm{x})  |\phi_{i}\right\rangle = \frac{i}{\hbar} \left\langle \phi_{f} | \mathcal{P} \mathcal{F}(\mathrm{x}) - \mathcal{F}(\mathrm{x}) \mathcal{P}  |\phi_{i} \right\rangle\,.
\end{equation}
	   Finally, by assuming  that the observable is a spatial coordinate  $\mathcal{F}(\mathrm{x})=\mathrm{x}$ (Feynman used a similar demonstration to show the correspondence of the FPI with canonical QM, \cite{Feynman:1942us}), the above equation  for generic   initial and final Hilbert  states $ |\phi_{i}\rangle $ and $ |\phi_{f}\rangle$,  is nothing but the commutation relation of ordinary QM: 
	   $
[\mathrm{x},\mathcal{P}]=i  \hbar 
$ --- or more in general $[\mathcal{F},\mathcal{P}]=i  \hbar \partial_\mathrm{x} \mathcal{F}$. 
	   With this result we have also checked the correspondence with canonical QM. The commutation relations, as well as the Heisenberg uncertain relation \cite{Dolce:2009ce}, can be regarded as \emph{implicit} 
(and not \emph{imposed}) in this theory.  This is a consequence of  the intrinsically cyclic (undulatory) nature of  elementary particles \cite{Dolce:2009ce}
conjectured by de Broglie in 1924 \cite{Broglie:1924}, implicitly tested by 80 years of successes of QFT and indirectly observed in a recent experiment \cite{2008FoPh...38..659C}.

\subsection{Semi-classical interactions}

In this section we want to generalize our geometrodynamical description of interaction to all the possible harmonic modes of the cyclic field. In this way we will find a formal correspondence with the ordinary FPI of interacting systems 
(Feynman diagrams and perturbative calculations will be discussed elsewhere \footnote{ The Fourier coefficients $ \bar a = \alpha_{1}$ and $ \bar a^* = \alpha_{-1}/\sqrt{-1}$   in (\ref{field:exp:modes})  can be regarded as the annihilating and  creation operators.  
The  anti-foundamental ($n=-1$) modes can be associated to an anti-particle.      Similarly,  $a_{n}=\alpha_{n}/\sqrt{n}$ can be regarded as Virasoro operators.  
}).     

We have already seen that the classical propagation of a free cyclic field is described by a FPI  (\ref{periodic:path.integr:Oper:Fey}) written in terms of the Lagrangian of a corresponding free particle with mass $\bar M$ \cite{Dolce:2009ce}. On the other hand,  we have also seen that local transformations of variables induce \emph{internal transformations} of field solutions  can be used to describe  gauge interaction. Next we will formally extend this correspondence to the quantum limit. In particular we will see that the propagation of the cyclic bosonic field - with all its harmonics - is described by the usual  FPI and Scattering Matrix of an ordinary gauge interacting bosonic fields. The  Lagrangian in FPI (\ref{periodic:path.integr:Oper:Fey})  will turn out to be the usual Lagrangian of classical electrodynamics of bosonic particles. Here we assume again three spatial dimensions $\mathbf x$.  

Next we will extend our semi-classical description of quantum systems to electrodynamics and we will find a formal correspondence to the corresponding QFT.

\subsubsection{Bohr-Sommerfeld quantization}

 The assumption of PBCs in field theory in compact 4D can be regarded as the quantization condition.  It is easy to see that PBCs reproduce the Bohr-Sommerfeld (BS) quantization condition. Intuitively the BS condition  is a periodicity condition because it says that the only possible orbits of the system are those with  an integer number of cycles.

The correspondence  with BS is immediate  in the case of free fields, since it corresponds to a isochronous system $T'^\mu(x)=T'^\mu$ (the analogous of to the Galilean isochronism of the pendulum where the orbits at different energies  have the same periodicity). The PBCs applied to the free cyclic field (\ref{field:exp:modes}) give the following harmonic quantization condition of the phase  of the fields
  \begin{equation}
 \oint d x \cdot p_{n} = \int^{T^\mu} d x \cdot p_{n}  = T \cdot p_{n}  = n h \,. 
  \label{BS:flat:4d}
\end{equation}
This is nothing but the harmonic spectrum of the four-momentum obtained in (\ref{quant:4P:free}). 
The quantization of a free field through PBCs is thus immediate because both the  four-periodicity and four-momentum are homogeneous. 

The interacting cyclic field  $\Phi'(x)$  is a sum of eigenmodes with the same analytical  form of the fundamental mode (\ref{transf:field:solution}),  
\beq
\bar \Phi(x')=\sum_{n }\bar \mathcal{N}_{n}  e^{-\frac{i}{\hbar}\int^{x'^\mu}d x \cdot \bar p'_n(x)}\,.\label{transf:gen:field:solution}
\eeq
The quantized spectrum of the cyclic interacting field in $x=X$ is given by  the PBCs at $T'^{\mu}(X)$, \emph{i.e.} $\Phi'(X) \equiv \Phi'(X+ cT(X))$, which can be written as 
\begin{equation}
e^{-\frac{i}{h} \int^{T'^\mu(X)} d x \cdot p'_{n}(x)}=e^{-\frac{i}{h} \oint_{X} d x \cdot p'_{n}(x)}  =   e^{-i 2 \pi n}\,.\label{quant:con:cyclicfield}
\end{equation}
 Therefore the quantization condition in the interacting case is 
 \begin{equation}
\oint_{X} d x \cdot p'_{n}(x)  =  n h  \,,\label{int:sommerfeld}
\end{equation}
according to (\ref{eq:deform:4mom:generic:int}) and (\ref{BS:flat:4d}). 
This is nothing but  the  generalization of the BS quantization condition in 4D \footnote{The Morse coefficient of the ordinary BS quantization can be retrieved by
  assuming a global twist factor in the PBCs and it can be interpreted as the \emph{vev} (see also Gauge-Higgs unification and Hosotani mechanism). This aspect has been shortly discussed in \cite{Dolce:2009ce}.}.   
 
We now define a four-momentum operator  $\mathcal{P}_\mu = \{\mathcal H/c, - \mathcal P_{i} \}$, and a number operator $\hat{n} \left|\phi_{n}\right\rangle \equiv n\left|\phi_{n}\right\rangle $.   In the Hilbert space associated with this interacting system the non-homogeneous  operator $\mathcal{P}'_{\mu}(x)$ can be obtained from the homogeneous one similarly to  (\ref{eq:deform:4mom:generic:int}),  
\begin{equation}
\mathcal{P}_{\mu} \rightarrow \mathcal{P}'_{\mu}(x) = e_{\;\mu}^{a}(x)  \mathcal{P}_{a} \,.\label{four:mom:oper:int}
\end{equation}
In fact the quantization condition  (\ref{int:sommerfeld}) of an interacting cyclic system can be expressed as 
\begin{equation}
\oint_{X} d x \cdot \mathcal{P}'(x)  = \hat n h\,.  \label{BS:oper:P}
\end{equation}

Therefore  the quantization of interacting cyclic field through PBCs can be also regarded as a generalization of the familiar BS quantization and described in terms of Hilbert operators.  

\subsubsection{Dirac quantization and  Symmetry Breaking}

We now consider the specific case of gauge interaction. For the cyclic field solution $\Phi'(x)$  associated with the  \emph{minimal substitution} (\ref{minim:subst:gener}),  the quantization condition (\ref{quant:con:cyclicfield}) turns out to be 
 \begin{equation}
e^{-\frac{i}{h} \oint_{X} d x \cdot [p_{n} - e A_n(x)]}  =   e^{-i 2 \pi n}\,. \label{quant:cond:gauge}
\end{equation}

As it can be  seen from (\ref{BS:flat:4d}),  in the case of pure gauge orbits $A_\mu(x)=\partial_\mu \theta(x)$, this condition leads to a quantization condition for the Wilson loop 
 \begin{equation}
 \oint_{X} d x \cdot  e A_n(x)  = h{ n} \,.  \label{quant:cond:Dirac}
\end{equation} 
Similarly to the Hamiltonian and momentum operator 
 we  define the operator $ e \mathcal{A}$ such that 
 \begin{equation} e\mathcal{A}\left|\phi_{n}\right\rangle \equiv eA_{n}\left|\phi_{n}\right\rangle\,. \label{def:A:oper} \end{equation}
In this case the assumption of PBCs  reproduces formally  quantization condition of a Dirac string \footnote{To the \protect \emph {l.h.s.} of this equation can be added a factor $1/2$ by  noticing that, according to the inner product (\ref {inner:prod}), only the
  modulo of the field has a physical meaning. In analogy with the XD formalism
  it arises naturally in the orbifold notation $s \in \protect \mathbb
  S^1/\protect \mathcal Z_2$. 
} \footnote{This condition can also be regarded as a quantization for the electric charge.
The  PBCs allows the possibility to use  Noether's theorem  to directly describe the quantized variables of the theory.  This idea will be expanded in future works.}.

\begin{equation}
 \oint_{X} d x \cdot  e \mathcal{A}(x)  =  
 \hat{n} h \, . \label{gauge:dirac:quant}
\end{equation}

This confirms the result  of the related paper \cite{Dolce:SuperC} where the Dirac quantization condition (\ref{gauge:dirac:quant})  has been obtained  in an indirect way just by assuming time periodicity for an abelian gauge field. It has been used to interpret phenomenological  aspects of superconductivity such as the quantization of the magnetic flux, the penetration length, the Meissner effect, the Josephon effect.  In this description, according to  \cite{Weinberg:1996kr,Dolce:SuperC},  superconductivity is a phenomenological consequence of the breaking of the electromagnetic gauge invariance associated with $\mathbb S^{1} \rightarrow \mathcal Z_{2}$.
 In fact,  in a pure gauge, because of the PBCs on the matter field $\Phi$, the phase of the gauge transformation (\ref{phase:transf:sec}) is periodic and defined modulo factor $h n$. This can be  seen also  from (\ref{gauge:dirac:quant}).  Thus the ``Goldstone'' $\theta$ of the related  gauge transformation can vary only by finite amounts and the electromagnetic gauge invariance  is broken  --- without involving a \emph{vev}. Notice however that this breaking of the gauge invariance  is a quantum effect  since in the classical limit $\hbar \rightarrow 0$ the quantization condition (\ref{gauge:dirac:quant})  yields a non quantized spectrum.     
Thus, field theory in compact 4D not only provides a geometrodynamical description of the gauge invariance; PBCs provides also a mechanism of gauge symmetry breaking with interesting analogy with the ones typically used in XD Higgsless and Gauge-Higgs-Unification models.  \footnote{ In the Hosotani  mechanism \cite{Hosotani1983193,Hall:2001tn},  the shift of the mass eigenvalue  of the KK fundamental mode  is $\bar M'(y)  = \bar M - g_{5} \bar A_{5}(y)$ where $\bar A_{5}(y)$ and $g_{5}$ are the fifth component and the coupling of an gauge field with an XD $y$. Under the dualism with XD studied \cite{Dolce:AdSCFT}  this is corresponds with  the \emph{minimal substitution} $\bar p'_{\mu}(x) = \bar p_{\mu} - e \bar A_{\mu}(x)$.   On the other hand, in complete parallelism with our 4D description, a Hilbert space can also be introduced to describe the KK mode of an XD field. The Wilson line of the   $A_{5}(y)$ can be written as an operator which takes integer number, similarly to  (\ref{gauge:dirac:quant}).  This corresponds to generalize the Hosotani mechanism from to winding number $n=1$ to all the possible $n$. As a results we find  a deformation of the whole KK tower, mode by mode,  which  can be equivalently described as induced by a corresponding deformation of the XD. 
}

As discussed in sec.(\ref{Non-abelian-case}), the generalization of the quantization condition (\ref{gauge:dirac:quant}) to the case of a non-abelian gauge  $SU_L(2) \otimes U_Y(1)$ is obtained through the following substitution  $e \mathcal{A}(x) \rightarrow g  \mathcal{W}(x) + g'  \mathcal{Y}(x)$.  
According to \cite{Csaki:2010cs}, the resulting quantization condition for the neutral component
\beq
 \oint_{X} d x \cdot \left[ g  \mathcal{W}(x) + g'  \mathcal{Y}(x) \right] =  
 \hat n h
\eeq
could provide a realistic electroweak symmetry breaking mechanism  which can be interpreted as induced by monopole condensations.  

\subsubsection{Scattering Matrix}

The Hilbert formalism (\ref{Schrod:repr:free})  used to describe the evolution of a free field $\Phi(x)$ can be easily extended to  interacting fields  $\Phi'(x')$. 

The transformation under  local abelian (polarized) isometries reproduces ordinary gauge interaction for the fundamental mode. Similarly to  (\ref{transf:field:solution:gauge}),  the corresponding  transformation of  a generic mode $\phi$  of a free cyclic field   to an corresponding mode $\phi'$ of a gauge interacting cyclic field   is described by the following internal transformation
\begin{equation}
\phi'(x) =  e^{\frac{i e}{\hbar} \int^{x^\mu} d x \cdot  A (x)} \phi (x)\,. 
\end{equation}
This describes the modulation of periodicity (\emph{tuning}) of the generic mode $\phi'$ with respect the free mode $\phi$, as a function of the generic gauge mode $A_\mu(x)$.  

Thus, from the definitions of the Hilbert operator $\mathcal A$ in (\ref{def:A:oper}),  we find that the \emph{internal transformation} of the cyclic field corresponds to pass  from the Schr\"odinger representation to the interaction  representation of perturbation theory. In fact we find 
\beq
|\phi'(x)\rangle =  e^{\frac{i }{\hbar} \int^{x^\mu} d x \cdot e \mathcal A} | \phi (x) \rangle = \sum_n  \alpha_n e^{\frac{i }{\hbar} \int^{x^\mu} d x \cdot eA_{n}} | \phi_n (x) \rangle\,. 
\eeq
Now we define a \emph{tuning} operator, Hilbert analogous of the \emph{parallel transport},  as 
\begin{equation}
\mathsf{S}(x)=e^{\frac{i}{\hbar} \int^{x^\mu} d^4 x \cdot e \mathcal{A} }\,.\label{scatt:matrix}
\end{equation}
This is nothing but the ordinary \emph{scattering matrix} of ordinary perturbation theory.
In fact, it turns out to define  formally  the interaction term $\mathcal L_{int} (x)$ of the ordinary Lagrangian of classical electrodynamics
\beq
e \int^{x^\mu} d x^\mu \mathcal A_\mu(x) 
=  e \int_{\tau_i^\mu}^{\tau_f^\mu} d \tau \mathcal A_\mu(x) \mathcal{J}^\mu(x) =  \int_{x_i}^{x_f} d^4 x  \mathcal{L}_{int} (x)\,.\label{L:int:classEM}
\eeq
In writing this equation we have explicitly written  the integration region as $x^\mu=x_f^\mu - x_i^\mu$ (modulo periods) and  we have defined the current $\mathcal{J}^\mu = d x^\mu / d \tau$. 

From a formal point of view, such a  representation of interaction for a cyclic field as modulation of periodicity actually matches the ordinary interaction representation written in terms of the \emph{scattering matrix} of ordinary QFT
\begin{equation}
|\phi'(x)\rangle = \mathsf{S}(x) | \phi (x) \rangle \,. 
\end{equation}
It is important to note that, as in  ordinary QFT, this interaction representation  is formally sufficient to describe QED  in terms of the Feynman diagrams. 

\subsubsection{Feynman Path Integral and Scalar QED}\label{FPI-QED}

With this formalism at hand it is finally possible to describe the evolution of an cyclic field (including all its possible harmonics) under a given interaction scheme. We will find that its evolution will be formally described by the ordinary FPI of the corresponding quantum interacting system. In the specific case of a cyclic field transforming under  abelian polarized local isometries, the result will be the ordinary FPI of QED (for bosons).

The description of the FPI given in (\ref{periodic:path.integr:Oper:Fey}), being written  in terms of elementary space-time evolutions,  can be easily generalized  to the non-interacting case. 
In  case of interactions  the  four-momentum operator $ \mathcal{P}'_{\mu}$ of (\ref{four:mom:oper:int}) is non-homogeneous.  Nevertheless the space-time evolutions are Markovian even in case of interactions . This can be seen for instance by considering  the scattering matrix (\ref{scatt:matrix}). In fact the time evolution in the interaction representation is described by the exponential operator $\mathsf{S}(t)=e^{-\frac{i}{\hbar} \int^{t}   \mathcal{H}_{int}(t)  d t }$. 
Similarly to the evolution of a free cyclic field (\ref{finite:time:evol}),  the contribution associated to  interactions is Markovian $\mathsf{S}(t+ d t)=e^{-\frac{i}{\hbar}    \mathcal{H}_{int}(t) dt }$.
Hence, in case of interactions the resulting  evolution  is Markovian: $\mathcal{U}'(t+dt;t)=e^{-\frac{i}{\hbar}    \mathcal{H}'(t) d t }$.   The generalization of the elementary space-time evolutions (\ref{elemet:phase:space}) of an interacting cyclic field in terms of the non-homogeneous four-momentum operator $\mathcal{P}'_{\mu} (x)$ is 
\begin{equation}
\mathcal U'(\mathbf{x}_{m+1},t_{m+1};\mathbf x_{m},t_{m}) 
  =  \left\langle \phi\right|e^{-\frac{i}{\hbar}(\mathcal{H}'\Delta t_{m}-\mathcal{P}'_{i}\Delta \mathbf{x}^{i}_{m})}\left|\phi\right\rangle 
   \,.\label{elemet:phase:space:int}
\end{equation}

 Another difference with respect to the free case is that interaction deforms point by point the  completeness relation. That is, the integration volume $V_\mathrm{x}$ of the inner  product (\ref{inner:prod}) varies with  $\mathbf{x}=\mathbf{X}$: that is,  $\int_{V'_\mathbf{x}(\mathbf{X})} {d \mathbf{x}'(\mathbf{X})}/{V'_\mathbf{x}(\mathbf{X})}$ (the number of period $N'$ remains fixed). The Markovian evolution of our cyclic system allows us to write the the evolution in terms of elementary evolutions (\ref{elemet:phase:space:int}). This guarantees the possibility to use a different inner product at  every point $\mathbf x= \mathbf x_{m}$ of (\ref{mark:element:PI}).  Moreover, in order to avoid  a different integration volume $\lambda'_\mathbf{x}(\mathbf{X})$ in every integration point, the integration region of the inner-product can be extended to a very large or infinite volume $V'_\mathbf{x}(\mathbf{X})$ (large or infinite number of periods $N'$), much bigger than the (finite) interaction region $\mathcal I$. In this way the volume $V'_{\mathbf{x}}(\mathbf{X})$, as well as  the normalization of the field,   is overall not affected by the local deformations: $V'_{\mathbf{x}}(\mathbf{X}) \cong V_{\mathbf{x}}$.    Thus the correct mathematical tool to represent this non trivial evolution is actually the integral product $\int_{V_{\mathbf{x}}} {\mathcal{D}\mathbf{x}}$.   

At this point, by following the same generic demonstration used in (\ref{periodic:path.integr:Oper:Fey}) (plugging locally the completeness relation in the elementary Markovian evolutions),  we find  formally the ordinary FPI in phase-space of an interacting particle
\begin{equation}
\mathcal{Z}=\lim_{N\rightarrow\infty}\int_{V_{\mathrm{x}}} \!\!\left(\prod_{m=1}^{N-1}{d\mathrm{x}_{m}}\!\right)\!\prod_{m=0}^{N-1}\!\left\langle \phi\right|e^{-\frac{i}{\hbar}(\mathcal{H}'\Delta t_{m}-\mathcal{P}'_{i}\Delta \mathrm{x}^{i}_{m})}\left|\phi\right\rangle\,. \label{periodic:path.integr:Oper:Fey:QED}
\end{equation}

Similarly to  (\ref{action:operator:generic}),   it is possible to define a Lagrangian $  {L}'_{cl}  = \mathcal{P}'_{i}\dot \mathrm{x}^{i} -  \mathcal{H}'$. 
Since $\mathcal{P}'_{\mu}$ in (\ref{four:mom:oper:int}) transforms as the four-momentum $\bar p'_{\mu}$ of the corresponding classical particle  (\ref{eq:deform:4mom:generic:int}), this lagrangian defines  formally the action ${\mathcal{S}}'_{cl}(t_{f},t_{i})\equiv\int_{t_{i}}^{t_{f}}dt {L}'_{cl} $ of the corresponding interacting classical particle --- written in terms of operators. 
Therefore the classical evolution of an interacting cyclic field is formally described by the ordinary  FPI in configuration space associated with the interaction scheme,
\begin{equation}
\mathcal{Z}=\int_{V_{\mathbf{x}}} {\mathcal{D}\mathbf{x}}e^{\frac{i}{\hbar}\mathcal{S'}_{cl}(t_{f},t_{i})}\,.\label{eq:Feynman:Path:Integral:QED}
\end{equation}

Finally, from the analysis of the scattering matrix (\ref{L:int:classEM}) we find  that in the approximation of the local  isometries investigated in this paper, the resulting evolution   is formally given by the ordinary  FPI of scalar QED. In fact the resulting Lagrangian associated with the action $\mathcal{S'}$ in the exponential is formally the classical Lagrangian of a  charged bosonic particle interacting electromagnetically:  
$\mathcal L'_{cl} = \mathcal L_{cl} +  \mathcal L_{int}$. 
  With this formal correspondence to the ordinary FPI formulation  at hand  we can invoke Feynman's  saying that  ``the same equations have the same solutions'' \cite{FeynmanLecVI}. Hence the assumption of intrinsic periodicity can in principle be used for a geometrodynamical  semi-classical description QED.
  
  This formal correspondence must be however tested in explicit computations of QED observables. This will be the subject of a dedicated paper. Nevertheless we mention that recent studies seem to show the liability of these semi-classical computations. We may note that light-front quantization, a semiclassical theory which similarly to our theory uses the assumption of PBCs as quantization condition, shows that it is actually possible to reproduce semi-classically the electron anomalous magnetic momentum in term of harmonics expansion of the fields \cite{Zhao:2011ct}.  Similar results pointing in the same direction are the computation of quantum behavior through the AdS/CFT, see discussion in sec.(\ref{AdSCFT}) and \cite{Dolce:AdSCFT}, and the calculation of Feynman diagrams in Twistor Theory \cite{ArkaniHamed:2009si} or other integrable theories.

\subsection{Further motivations}

Here we present a digression about the physical meaning of the assumption of intrinsic periodicity for elementary quantum systems. The motivations go  beyond the de Broglie periodic phenomenon and involve interesting aspects of modern physics, as described in detail in \cite{Dolce:2009ce,Dolce:2009cev4,Dolce:2010ij,Dolce:QTRF5,Dolce:Dice,Dolce:FQXi,Dolce:Cyclic,Dolce:AdSCFT,Dolce:SuperC} and summarized here.  

We may consider  the recent attempts to interpret QM as an emerging theory,  such as  the 't Hooft determinism \cite{'tHooft:2001ar} and  the stroboscopic quantization \cite{Elze:2003tb}. According to 't Hooft \cite{'tHooft:2001ar}, there is a ``close relationship between the quantum harmonic oscillator and a particle moving on a circle'', both with extremely fast periodicity $T_{t}$. Our field theory in compact 4D can be intuitively derived by noticing that, as well known, the quantum harmonic oscillator is the basic element of  ordinary second quantized KG fields. A cyclic field can be intuitively derived from the 't Hooft determinism (in the continuos limit of the lattice used by 't Hooft) by considering that the characteristic periodicity $T_{t}$ varies in a relativistic way, as described in sec.(\ref{Compact4D:formalism}). This cyclic behavior of a ``particle on a circle'' of the 't Hooft determinism has motivated the ``stroboscopic quantization'' \cite{Elze:2003tb}. In this case we explicitly find the idea of dimensions compactified in a torus and, even more interesting, the fact that the ``ticks'' resulting from ergodic dynamics yield  an effective  description of the arrow of time. Similarly, in our theory every instant in time can be characterized by a different combination of the phases of all the de Broglie clocks constituting an isolated system of elementary particles. This is similar to a calendar or a stopwatch which allows us to fix events in time in term of combinations of the phases of periods that we call years, months, days, hours, minutes, and so on. If the elementary cycles constituting our systems of periodic phenomena have irrational periods, the total evolution will result in an ergodic evolution; if we also allow interactions between elementary periodic phenomena, \emph{i.e.}  exchange energy or equivalently variation of periodicity, the resulting evolution of the non elementary system will be chaotic. Indeed, the   assumption of intrinsic periodicity, realized in terms of field theory in compact 4D with PBCs, has important motivations in the interpretation of the notion of time in physics \cite{Dolce:2009ce,Dolce:FQXi}. In particular, as Galileo taught us with the experiment of the pendulum in the Pisa dome, or as explicitly stated in the Einstein \cite{Einstein:1910} definition of relativistic clock, or according to the operative definition of a \emph{second} through the Cs-133 atom, time can be only defined by counting the number of periods of a phenomenon supposed to be periodic in order to guarantee the constancy of the unit of time. \emph{Thus every free elementary particle can be regarded as a reference clock} \footnote{In this way the local character of the relativistic time turns out to be enforced.}, the so-called ``de Broglie internal clock''. As we have seen,  the modulation of periodicity of these internal clocks can be used to describe interactions similarly to GR. 
We also mention the geometric quantization \cite{Gozzi:2006cv} which is an attempt to reproduce quantum behavior by introducing two grassmannian partners of the physical time. This could be  interesting for a possible semi-classical description of spinning particle typical of the \emph{zitterbewegung} models \cite{Casalbuoni:1976tz} (this brings elements of supersymmetry in the theory and  it could be investigated in a dedicated paper).    

The  time periodicity of the de Broglie internal clock is bounded by the inverse of the mass, $T_{\tau} \leq T_{\tau} = h / \bar M c^{2} $, \emph{i.e.} by the Compton wavelength divided by the seed of light  $T_{\tau} = \lambda_{s} / c$. In this way it is easy to see that these de Broglie intrinsic clocks of elementary particles  are typically extremely fast (except neutrinos). The heavier the particle, the faster the periodicity. As already mentioned, a light particle such as the electron has a periodicity faster than $10^{-20} s$ which means that for every ``tick'' of the Cs-133 atomic clock ($T_{Cs} \sim 10^{-10} s$) an electron does a number of cycles of the order of the age of the universe expressed in  years. Even with the modern time resolution  it is not yet possible to resolve such small time scales, though  the internal clock of the electron has been indirectly observed in a recent interference experiment \cite{1996FoPhL}. Thus the observation of such a fast de Broglie internal clock is similar to the observation of a ``clock under a stroboscopic light'', \cite{Elze:2003tb}. That is, at every observation the particle appears to be in an aleatory phase of its cyclic evolution and, similarly to a dice rolling too fast (de Broglie deterministic dice) with respect to our resolution in time, the outcomes can be only described in a statistical way \cite{Dolce:Dice}. Similarly to the deterministic models mentioned above, the results of the preview section show that the statistics associated to these cyclic behavior have formal correspondences to ordinary QM. They also suggest that the direct experimental exploration of microscopical time scales (smaller that $10^{-20} s$ in the case of the electron) is of primary interest in understanding the inner nature of the elementary systems\footnote{We can say that LHC is exploring indirectly time scale of the order of $10^{-27} s$ corresponding to energy scale of the order of the $TeV$.}.

Another motivation is given by the variational analysis of the BCs discussed in \cite{Dolce:2009ce} and mentioned in sec.(\ref{Compact4D:formalism}). Roughly speaking we may say that relativity fixes the differential structure of the continuous space-time of a theory without giving particular prescriptions about BCs, \cite{Dolce:2009ce,springerlink:10.1007/BF01889475}. The important requirement for the BCs is that they must fulfill the variational principle. As already discussed in this paper, the role of the BCs is marginal in ordinary QFT.  On the other hand, BCs have played an important role since the earliest days of QM, according to de Broglie, Bohr, Sommerfeld, et.al..   
 We have also noticed that the non-quantum limit of a massive cyclic field corresponds to the limit where only the fundamental mode $\bar \Phi(x)$ is exited, see \cite{Dolce:2009ce} and \cite{Dolce:2009cev4} for more detail. This also is the non-relativistic limit of a massive particle ($|\bar{\mathbf p}| \ll \bar M c $). Thus,  the classical particle description, corresponding to the limit of large mass ($\bar M \rightarrow \infty$) or equivalently the spatial momentum to zero ($\bar{\mathbf p} \rightarrow 0$), the cyclic field reduces to $\bar \Phi(x) \sim \exp[-i \frac{\bar M c^{2}}{\hbar} t + i \frac{\bar M}{\hbar} \frac{\mathbf x^{2}}{2 t}]$, see \cite{Dolce:2009ce}. Neglecting the de Broglie rest clock represented by the first term, the modu\-lo square of a massive cyclic field is a distribution centered along the path of the corresponding classical particle. Its width is of the order or smaller of the Compton wavelength $\lambda_{s}$, as can be easily shown by performing an explicit plot  \cite{Dolce:2009cev4}. Thus in the classical limit ($\lambda_{s} \rightarrow 0$)   this distribution reproduces the ordinary non-relativistic limit of the FPI, \emph{i.e.} a Dirac delta distribution. In this limit the spatial compactification lengths tend to infinity ($\bar{\mathbf p} \rightarrow 0$) whereas the compactification length along the time dimensions tends to zero ($\bar M \rightarrow \infty$). Indeed a classical particle turns out to be actually described by a point like distribution in $\mathbb R^{3}$. Similar arguments can be used to interpret other interesting aspects of the wave-particle duality of QM. The assumption of periodic phenomenon enforces the wave-particle dualism, giving rise to implicit commutations relations and Heisenberg uncertain relations, \cite{Dolce:2009ce}. A massless field, \emph{i.e.} a field leaving in the light-cone ($d s = 0$), is always relativistic. Since its Compton wavelength is infinite we say that the rest de Broglie clock of a massless field, such as the EM field, is \emph{frozen}. Thus the energy spectrum of a massless cyclic field can be approximated to a continuous in the IR region where the compactification length tend to infinity ($T_t \rightarrow \infty$) and the PBCs can be neglected. In the UV limit however the PBCs are important because the compactification length are very small ($T_t \rightarrow 0$). Therefore, the quantized nature of the energy spectrum becomes manifest avoiding the UV catastrophe of the black-body radiation.  Indeed  at high frequencies the field theory has an effective corpuscular description, \cite{Dolce:2009ce}.

A further conceptual motivation to use BCs as quantization condition is that we have the remarkable property that QM emerges without involving any (local) hidden variable in the theory. Since the hypothesis of existence of local hidden variable is not realized, the Bells's theorem can not be applied to our theory. The assumption of intrinsic periodicity introduces an element of non locality which, however, can be regarded as consistent with SR since the periodicity varies in a relativistic way. Thus the theory can in principle violate the Bells's inequality (if we try to adapt the Bell's theorem to our case, \emph{i.e.} to evaluate the expectation values of an observable in the Hilbert space described above, we find again a formal parallelism with ordinary QM).  For this reason we speak about --- mathematically --- deterministic theories \cite{Dolce:2009ce}.   

 We have seen that the assumption of intrinsic periodicity of elementary systems provides a semi-classical description of scalar QED. The study of the limit where quantum corrections become relevant for gravitational interaction is beyond the scope of this paper. In particular quantum gravity represents another important subject where to test the consistence and the validity of the theory. Some more detail about this point is given in \cite{Dolce:2009ce,Dolce:FQXi}.

\section{Conclusions}

 Field theory in compact 4D  represents a natural realization of the ``periodic phenomenon'' associated to every elementary particle, as conjectured by de Broglie in 1924 \cite{Broglie:1925,Broglie:1924},  at the base of the wave-particle duality, implicitly tested by 80 years of QFT and indirectly observed in a recent experiment \cite{2008FoPh...38..659C} (Schr\"odinger used a similar assumption in his \emph{zitterbewegung} model of the electron). 
 
 In this formalism the kinematical information  of an interaction scheme  is encoded in the relativistic geometrodynamics of the boundary of the theory --- in a sort of holographic description.
The resulting description has shown remarkable relationships with the following  fundamental approaches to interactions:
\begin{itemize}
\item the  approach typical of classical-relativistic mechanics in which  interaction is described in terms of retarded and local variations of four-momentum;
\item the approach  typically used in QM to describe systems  in generic potentials in terms waves and BCs. The dynamical boundary of the theory reproduces the retarded and local modulation of de Broglie four-periodicity and thus the local and retarded variation of four-momentum of an interacting quantum system;
\item  in GR,  gravitational interaction can be described as local modulations of periodicity of reference clocks encoded in corresponding  deformations of the underlying space-time coordinates.
Similarly, in our description,  local modulations of the four-periodicity are described as deformations of the compact 4D. 
\item the  typical approach to interaction of gauge theory is obtained because the variations  of field solution associated with the variations of the boundary of the theory defines an \emph{internal transformations} which, in the approximation of the local isometries described in the paper, formally matches classical electrodynamics. 
\end{itemize}
Remarkably, we have found that  \emph{gauge interaction can be derived from the invariance of the theory under  local transformations of variables}  as gravitational interaction can be derived by requiring invariance  under diffeomorphisms. Gauge symmetries are related to space-time symmetries. 
This cab be regarded as in the spirit of  Weyl's, Kaluza's and Wheeler's and  original proposal of a geometrodynamical description of gauge invariance.  

On the other hand the assumption of PBCs (or similar BCs such as anti-PBCs, N-BCs, D-BCs) provides a semi-classical quantization condition for  fields.  In fact the theory can be regarded as the full relativistic generalization of the quantization of a ``particle in a box''. 
This geometric quantization method, without introducing hidden-variables, \emph{reproduces formal correspondences between  well established quantization methods}:
\begin{itemize}
\item the correspondence to  canonical formulation of QM, arises from the fact that  a cyclic field is  naturally described by an Hilbert space, that it evolves according to the Schr\"odinger equation and its cyclic  variables implicitly satisfy commutation relations and Heisenberg uncertain relations;
\item the Feynman Path Integral is obtained as interference of the classical paths with different winding numbers associated with the intrinsically cyclic geometry of fields in compact 4D \cite{Dolce:2009ce,Dolce:2009cev4};
\item The Bohr-Sommerfeld quantization condition  is nothing but a periodicity condition. It simply states that the allowed orbits are those with an integer number of cycles (close orbits) and can be described in terms of PBCs;
\item the Dirac quantization condition is obtained as a result of the periodicity induced by the matter cyclic field on pure gauge transformations. 
\item the  Scattering Matrix naturally describes the modulations of periodicity between a free cyclic field (Schr\"odinger representation) and an interacting cyclic field (interaction representêation).   

\end{itemize}

Remarkably, \emph{field  theory in cyclic 4D, without any further assumption than intrinsic periodicity, provides the possibility of a geometrodynamical and semi-classical description of scalar QED}. 

To the above list of well established quantization methods it should be added that, as shown in \cite{Dolce:AdSCFT} through  the dualism of cyclic fields  to  XD massless fields,  the theory yields interesting analogies with the classical XD geometry to quantum behavior correspondence typical of AdS/CFT. 
The dualism to XD theory will be used in future papers to investigate the geometrodynamical description of gauge invariance in terms of Kaluza's original proposal. 

 \section*{Acknowledgements} 
 I would like to thank M. Neubert, M. Reuter, E. Manrique and T. Gherghetta  for their interest in new ideas; N. Liu  for her help in writing; and the reviewers for their valuable work.
  This paper is part of the project ``\emph{Compact Time and Determinism}'' and it has been presented at  \texttt{QTS7} Aug 2011, Prague, and \texttt{FFP12} Nov 2011, Udine. 

\providecommand{\href}[2]{#2}\begingroup\raggedright

\end{document}